\documentclass[aps,twocolumn,prb,showpacs,tightenlines]{revtex4}
\usepackage{amssymb}
\usepackage{epsfig}
\usepackage{amsbsy}
\usepackage{amsmath}
\usepackage{graphicx}

\begin{document}

\title{Lehmann-Symanzik-Zimmermann Reduction Approach to \\
Multi-Photon Scattering in Coupled-Resonator Arrays}
\author{T. Shi and C. P. Sun}
\affiliation{Institute of Theoretical Physics, Chinese Academy of Sciences, Beijing,
100190, China \\
}

\begin{abstract}
We present a quantum field theoretical approach based on the
Lehmann-Symanzik-Zimmermann reduction for the multi-photon scattering
process in a nano-architecture consisting of the coupled-resonator arrays
(CRA), which are also coupled to some artificial atoms as a controlling
quantum node. By making use of this approach, we find the bound states of
single photon for an elementary unit, the T-type CRA, and explicitly obtain
its multi-photon scattering S-matrix in various situations. We also use this
method to calculate the multi-photon S-matrices for the more complex quantum
network constructed with main T-type CRA's, such as a H-type CRA waveguide.
\end{abstract}

\maketitle

\section{Introduction}

In order to realize all-optical quantum devices~\cite{NP07,Fan05}, it is
very crucial to explore the physical mechanism for the single photon
generation, transport and one shot detection \textit{et\thinspace al.}%
\thinspace . Thus, we need a comprehensive understanding of the fundamental
processes of coherent photonic scattering in the solid state based confined
systems, such as the photonic crystal with artificial band gaps.
Essentially, by coupling the system with an extra two-level system (TLS) to
form a hybrid system, for controllably transport of photons the basic
element is a quantum node or quantum switch~\cite{Sun}. It was abstracted as
the so-called photon transistor most recently~\cite{NP07, Zhou}. For quantum
information processing, such quantum node can coherently control the quantum
state transfer in some quantum network~\cite{ST,YS1,YS2}. Actually, to
manipulate the coherent transport of photons, the quantum node is tunable so
that it can behave either as a perfect mirror totally reflecting photons, or
as an ideal transparent medium allowing photons to pass throughly.
Theoretically, the quantum node for single-photon in all-optical
architectures was extensively studied in one dimensional waveguide by making
use of the standard scattering approaches~\cite%
{Fan05,FanPRL,FanPRA,FanOpt,Zhou2,FMH,Nori2}.

The photonic quantum node is usually modelled as a localized TLS~\cite{Dicke}%
, which can be implemented as an artificial atom, coupled to the photons
transported in the coupled-resonator arrays (CRA)~\cite{Nori}. The atomic
parameters, e.g., the energy level spacing, are tunable to control the
propagation of photons. Recently, based on this theoretical model and its
generalizations, the photon transport in the CRA systems has been studied
for different purposes~\cite{Zhou,Xu,FanPRL,FanPRA}. Here, we would like to
point out two remarkable issues: (a) if only one photon allowed to transport
in the CRA with the artificial atom prepared at its ground state, the hybrid
system can be described by a simple model with a single state coupled to a
continuum, which is referred to the Anderson-Fano-Lee model~\cite{F,A,L}
with single excitation; (b) the CRA can be regarded as the waveguide with
nonlinear dispersion relation. It can be linearized in the high frequency
regime in which photon momentum $k\sim \pm \pi /2a$ with lattice spacing $a$%
, to approximate the linear dispersion relation for the conventional
waveguide.

In the recent study~\cite{Zhou}, the single photon transmission and
reflection coefficients were calculated for the incident photon with any
energy in the T-type structure (a quantum node coupled to a CRA, see Fig.
1), which demonstrate the novel lineshapes beyond the Breit-Wigner~\cite%
{Landau} and the Fano lineshapes~\cite{F}. This kind of investigation was
carried out only for the case with single photon. Also, there are only quite
a few researches on the two photon case, for which we mention an elegant
theoretical approach for two photon scattering~\cite{FanPRL,FanPRA} based on
the Bethe-ansatz~\cite%
{Bethe,Yang,Batchelor,KondoAndrei,Wiegmann1,Wiegmann2,Wiegmann3}. Moreover,
we have to say that it sounds very difficult to prepare the system only with
one photon or two photons exactly, thus these previous studies need to be
improved for multi-photon processes oriented by practical application.

Actually, the study for multi-photon transport is very important to realize
a practical all-optical devices. However, these subtle approaches for single
and two photon cases mentioned above~\cite{Zhou,Xu,FanPRL,FanPRA}, such as
the discrete coordinate scattering theory and the Bethe-ansatz technique
with fixed scatterers, are not feasibly generalized for the realistic
multi-photon processes, even for two-photon or three-photon processes.
Therefore, our present systematical approach based on quantum field theory
is significant since it is obviously feasible and intrinsically natural for
the generalization to multi-photon scattering processes.

In this paper, we utilize the Lehmann-Symanzik-Zimmermann (LSZ) reduction~%
\cite{LSZ} in quantum field theory to investigate the multi-photon transport
in the complex CRA with some two- level scatterers. This method has been
used to study the single electron inelastic scattering in Anderson model and
Kondo model~\cite{Andrei1,Andrei2}. Here, we deal with the multi-photon
scattering problem by studying the out-state of the scattered photons for an
arbitrary state of incident photons. In a middle stages, we calculate the
multi-photon scattering matrix ($S$-matrix) in details. With the
diagrammatic analysis, we find that the basic element of the $S$-matrix is a
connected transfer matrix ($T$-matrix), which can be obtained from the
well-known LSZ reduction formula about the photonic Green's function. From
the explicitly achieved expressions of photonic out-states, we analysis in
details quantum statistical characters of photon transmission in the
situation with many photons. We find that, in the tight binding CRA of
T-type, there exist the single photon bound states. As a test, two photon
transport in the T-type waveguide is re-considered, and our obtained results
accord with the recent woks~\cite{FanPRL,FanPRA} using the Bethe-ansatz,
which verifies the results based on the LSZ reduction approach are valid. As
the development, the three photon scattering is studied, and the outgoing
states of the three photon are given by this approach. Our present
investigation mainly based on these results, can be regarded as a
substantial development for its particular emphasis on the multi-photon
scattering.

In practice, the T-type photonic element we mentioned above is the basic
block to constitute a complex quantum network coherently transferring
photons in a controllable fashion. A slightly complicated illustration of
such architecture is the CRA waveguide of H-type. In this paper, we also
study two photon scattering processes in the H-type in details.

The paper is organized as follows: in Sec.~II and~III, we model our hybrid
system for multi-photon transport and present the scattering matrix based on
the LSZ reduction approach; in Sec.~IV, we show that there exist the single
photon bound states in the tight binding T-type CRA; in Sec.~V, we study the
multi-photon transport in the T-type waveguide; in Sec.~VI, we discuss the
two photon transport in the H-type waveguide; in Sec.~VII, the results are
summarized with some remarks.

\section{Scattering model for the hybrid system}

\subsection{Model setup}

\begin{figure}[tbp]
\includegraphics[bb=51 171 498 657, width=6 cm, clip]{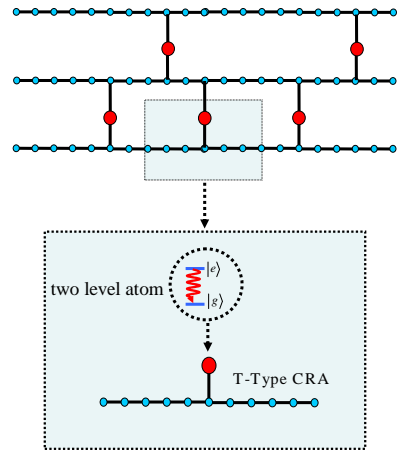}
\caption{(Color online) The schematic for the complex CRA: The T-type
structure with CRA coupled to impurity (maybe a two-level scatterer)is the
basic element to construct the more complicated architecture of quantum
network. The red circles denote the two level impurity, while the blue dots
denote the photonic coupled resonators.}
\label{fig1}
\end{figure}

In this subsection, we model the T-type CRA by the two-level atom coupled to
photons inside CRA\ illustrated in Fig.~\ref{fig1}. The model Hamiltonian
reads%
\begin{eqnarray}
H_{T} &=&\Omega \left\vert e\right\rangle \!\left\langle e\right\vert
+\sum_{i}[\omega _{0}a_{i}^{\dagger }a_{i}-J(a_{i}^{\dagger }a_{i+1}+\text{%
\textrm{H.c.}})]  \notag \\
&&+V\sum_{i}\delta _{i0}(a_{i}^{\dagger }\sigma ^{-}+h.c.)\,,  \label{Hx}
\end{eqnarray}%
where the operator $\sigma ^{-}=\left\vert g\right\rangle \left\langle
e\right\vert $ denotes the flip from the atomic ground state $\left\vert
g\right\rangle \ $to the excited state $\left\vert e\right\rangle $ with the
energy level spacing $\Omega $. Here, $J$ is the\ hopping constant\
characterizing the inter-cavity coupling in the tight-binding approximation;
$a_{i}\,$($a_{i}^{\dagger }$) is the annihilation$\,$(creation) operator for
the photonic single mode with eigen-frequency $\omega _{0}$ in the $i$-th
cavity; $V$ is the hybridization constant of the localized atom-photon in
the $0$-th site of CRA.

In the momentum space ($k$-space), the Hamiltonian~(\ref{Hx}) is
re-expressed as%
\begin{eqnarray}
H_{T} &=&\Omega \left\vert e\right\rangle \!\left\langle e\right\vert
+\sum_{k}\varepsilon _{k}a_{k}^{\dagger }a_{k}  \notag \\
&&+\frac{V}{\sqrt{L}}\sum_{k}(a_{k}^{\dagger }\sigma ^{-}+\text{\textrm{H.c.}%
})\,,  \label{HT1}
\end{eqnarray}%
with the photonic dispersion relation%
\begin{equation}
\varepsilon _{k}=\omega _{0}-2J\cos k\,,
\end{equation}%
in CRA\ of length $L$. Here, we choose the cavity spacing $a=1$. In the high
energy limits $k\sim \pm \pi /2$ and $\omega _{0}=\pi J$, the above
dispersion relation is linearized as $\varepsilon _{k}\sim \left\vert
k\right\vert $, which is the same as that in the conventional waveguide.
Thus, we can use CRA to simulate the conventional waveguide in the high
frequency limits.
As the basic element of the complex quantum network as illustrated in Fig.~%
\ref{fig1}, the two- level atom plays the role of the photon transistor to
control the photon transmission.

It is obvious that, when confined within the single excitation subspace, our
model is the same as the models by Anderson, Fano and Lee. This observation
motivates us to consider how to use various approaches developed previously
for the models by Anderson, Fano and Lee, to deal with the coherent
processes of our system, especially with multi-photons. To this end, we will
compare our model (\ref{HT1}) with the Anderson, Fano and Lee model as
follows.

\subsection{Relation to Anderson model}

By neglecting the Coulomb interaction, the Anderson model is the same as the
Fano model, so we discuss the similarities and differences between the
model~(\ref{HT1}) and Anderson model. The corresponding Hamiltonian reads%
\begin{equation}
H_{A}=\sum_{k,\sigma }\varepsilon _{k}c_{k\sigma }^{\dagger }c_{k\sigma
}+\varepsilon _{d}f_{\sigma }^{\dagger }f_{\sigma }+H_{V}+H_{U}\,,
\label{Anderson}
\end{equation}%
where $c_{k\sigma }\,$($c_{k\sigma }^{\dagger }$) is the annihilation
(creation) operator of the conductive electron with dispersion relation $%
\varepsilon _{k}$ and spin $\sigma $. $f_{\sigma }\,$($f_{\sigma }^{\dagger
} $) is the annihilation$\,$(creation) operator of the impurity $f$-electron
with energy $\varepsilon _{d}$ and spin $\sigma $. The Coulomb repulsive
interaction of the impurity $f$-electrons is described by the Hubbard term $%
H_{U}=Uf_{\uparrow }^{\dagger }f_{\uparrow }f_{\downarrow }^{\dagger
}f_{\downarrow }$. The hybridization of a conductive electron with the
localized $f$-electron is depicted by the mixing term%
\begin{equation}
H_{V}=V\sum_{k,\sigma }(c_{k\sigma }^{\dagger }f_{\sigma }+\text{\textrm{H.c.%
}})\,.
\end{equation}

Anderson even used this model~(\ref{Anderson}) to investigate the localized
magnetic state in the metal. By comparing the Hamiltonian~(\ref{Anderson})
with the Hamiltonian~(\ref{HT1}), we find that the atom and the photon in
the model~(\ref{HT1}) can be considered as the impurity $f$-electron and the
conductive electron in the Anderson model, respectively. But there still
exist some differences between the two models. First, the Hilbert space of
the atom in the model~(\ref{HT1}) is spanned by two basis states, $%
\left\vert e\right\rangle $ and $\left\vert g\right\rangle $, but the
Hilbert space of the impurity $f$-electron is spanned by four basis states, $%
\left\vert 0\right\rangle $, $f_{\uparrow }^{\dagger }\left\vert
0\right\rangle $, $f_{\downarrow }^{\dagger }\left\vert 0\right\rangle $ and
$f_{\uparrow }^{\dagger }f_{\downarrow }^{\dagger }\left\vert 0\right\rangle
$. Secondly, in the Anderson model the conductive electrons are fermions,
while in the model~(\ref{HT1}) photons are bosons. Thirdly, there is no
Hubbard term $H_{U}$ in the model~(\ref{HT1}), while in the Anderson model
if there is only one $f$-electron the Coulomb interaction vanishes, i.e.,
the Hubbard interaction $H_{U}$ does not play any role. In conclusion, the
two models are equivalent only in the single excitation case, i.e., one
conductive electron with no $f$-electron or one impurity $f$-electron with
no conductive electron in the Anderson model, and one photon with the atom
prepared at the ground state or no photon with the atom prepared at the
excited state in the model~(\ref{HT1}). Otherwise, in the multi-excitation
case, the two models~(\ref{HT1}) and~(\ref{Anderson}) are obviously not
equivalent.

\subsection{Relation to Lee model}

The Lee model~\cite{L} with the Hamiltonian%
\begin{eqnarray}
H_{L} &=&m_{V}\psi _{V}^{\dagger }\psi _{V}+m_{N}\psi _{N}^{\dagger }\psi
_{N}+\sum_{k}\omega _{k}A_{k}^{\dagger }A_{k}  \notag \\
&&+g\sum_{k}\frac{1}{\sqrt{2\omega _{k}V}}(\psi _{V}^{\dagger }\psi
_{N}A_{k}+\text{\textrm{H.c.}})\,,  \label{Lee}
\end{eqnarray}%
describes a reaction process%
\begin{equation}
V\rightleftarrows N+A_{k}\,,
\end{equation}%
with two heavy fermions ($V$, $N$) of masses $m_{V}$ and $m_{N}$. Here, the
relativistic boson $A_{k}$ with momentum $k$ and rest mass $\mu $\ possesses
the dispersion relation $\omega _{k}=\sqrt{k^{2}+\mu ^{2}}$, and $g$ is the
three body coupling constant for the scattering of the three kinds of
particles, $V$, $N$, and $A$. T. D. Lee used this exact solvable model to
study the necessity of renormalization in quantum field theory even without
the perturbation expansion.

By comparing Lee model with the model~(\ref{HT1}), we find that the heavy
fermions $V$ and $N$ can be regarded as the excited state and ground state
of the atom, respectively. And the relativistic boson $A_{k}$ can be
considered as the photon in the model~(\ref{HT1}). There also exist some
differences between the Lee model and the model~(\ref{HT1}). Firstly, in Lee
model, the renormalizations of the $V$-fermion mass $m_{V}$ and coupling
constant $g$ should be taken into account, and the Hamiltonian~(\ref{Lee})
is non-Hermitian. Thus we should investigate the unitarity of the $S$-matrix
carefully~\cite{Pauli,Lee-Wick}. Secondly, the Hilbert space of heavy
fermion is spanned by four basis states, i.e., $\left\vert 0\right\rangle $,
$\psi _{V}^{\dagger }\left\vert 0\right\rangle $, $\psi _{N}^{\dagger
}\left\vert 0\right\rangle $ and $\psi _{V}^{\dagger }\psi _{N}^{\dagger
}\left\vert 0\right\rangle $. We conclude that the two models~(\ref{HT1})
and~(\ref{Lee}) are equivalent only in the single excitation case, i.e., one
relativistic boson $A_{k}$ with one heavy fermion $N$ or one heavy fermion $%
V $ with no relativistic boson $A_{k}$ in Lee model, and one photon with the
atom prepared at the ground state or no photon with the atom prepared at the
excited state in the model~(\ref{HT1}).

In fact, it is turn out that the multi-particle scattering problems (such as
$N$-$\theta \theta $ scattering) in the Anderson and Lee models were both
successfully studied by the LSZ reduction approach~\cite%
{Andrei1,Andrei2,Maxon1,Maxon2}. In the following, we start with the model~(%
\ref{HT1}) to investigate the $S$-matrix of scattering photon in the hybrid
CRA systems. For the further discussion, we first give a general formalism
to study the multi-photon $S$-matrix in the complex CRA by the LSZ reduction
formalism.

\section{Lehmann-Symanzik-Zimmermann reduction for photon scattering in
general}

In this section, we first briefly summarize the main results of the LSZ
reduction approach in quantum field theory, and then we use them to give a
general formula for multi-photon $S$-matrix in the T-type CRA. For
investigating the multi-photon scattering, in the first subsection, we first
define the $n$-photon $S$-matrix and $2n$-point photonic Green's function,
which are explicitly described by the Feynman diagrams (Fig.~\ref{fig2}),
and utilized the functional integral to obtain the $2n$-point photonic
Green's function. In the second subsection, we use the LSZ reduction
approach to obtain a general form of the $n$-photon $S$-matrix element by
reducing the external legs (red curves in Fig.~\ref{fig2}) of the Green's
function.

\subsection{Main results of the LSZ reduction approach}

Because the LSZ reduction approach relates to the photonic Green's function
and $S$-matrix, we need to give their definitions for discussing the LSZ
reduction explicitly. The $2n$-point photonic Green's function%
\begin{eqnarray}
&&G_{p_{1},...,p_{n};k_{1},...,k_{n}}\!(t_{1}^{\prime },...,t_{n}^{\prime
};t_{1},...,t_{n})  \notag \\
&=&\left\langle \Theta \right\vert Ta_{p_{1}}\!(t_{1}^{\prime
})...a_{p_{n}}\!(t_{n}^{\prime })a_{k_{1}}^{\dagger
}\!(t_{1})...a_{k_{n}}^{\dagger }\!(t_{n})\left\vert \Theta \right\rangle \,,
\label{Green}
\end{eqnarray}%
is defined by the time ordering product $T$ of the photon creation
(annihilation) operator $a_{k}^{\dagger }$\thinspace ($a_{k}$) in the
Heisenberg picture. Here, $\Theta $ denotes the ground state of the
Hamiltonian $H$ of the system.

The $S$-matrix element \cite{Sakurai,Taylor}%
\begin{equation}
_{out}\left\langle f\left\vert i\right\rangle _{in}\right. =\left.
_{in}\left\langle f\right\vert S\left\vert i\right\rangle _{in}\right. \,.
\label{S}
\end{equation}%
of photons is defined through the overlap of incoming state $\left\vert
i\right\rangle _{in}$ and outgoing scattering state $\left\vert
f\right\rangle _{out}$, which are asymptotically free, and may contain
multi-photon excitation, i.e., they are multi-photon states. In our case,
the incoming state is%
\begin{equation}
\left\vert i\right\rangle _{in}=a_{k_{1}}^{\dagger }...a_{k_{n}}^{\dagger
}\left\vert 0\right\rangle \,.
\end{equation}%
In the present case, the $S$-matrix is given by%
\begin{equation}
S=T\exp [-i\int_{-\infty }^{+\infty }dt\,H_{int}\!(t)]\,,
\end{equation}%
with the atom-photon hybridization Hamiltonian%
\begin{equation}
H_{int}=V(a_{0}^{\dagger }\sigma ^{-}+\text{\textrm{H.c.}})\,,
\end{equation}%
in the interaction picture.

\begin{figure}[tbp]
\includegraphics[bb=62 220 595 684, width=8 cm, clip]{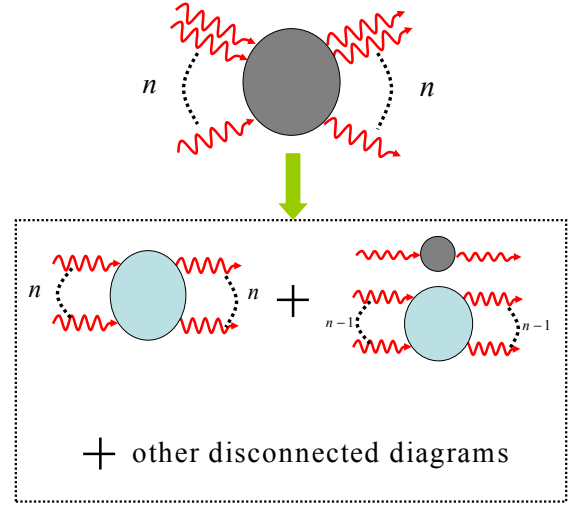}
\caption{(Color online) The Feynman diagrams for LSZ reduction: The red line
denotes the free photon propagator. The gray shade circles denote the $S$%
-matrix elements and the blue shade circles denote the $T$-matrix elements.
The Feynman diagram of $S$-matrix element is constructed by all kinds of the
disconnected diagrams. And each of the the disconnected diagrams contains
some connected diagrams.}
\label{fig2}
\end{figure}

Using the diagrammatic analysis, we find that the Feynman diagram of $S$%
-matrix element is constructed by summing up the contributions from all
kinds of disconnected diagrams as shown in Fig.~\ref{fig2}. Here, each of
these disconnected diagram is made up of some connected diagrams which
describe the $T$-matrix elements.
As the special cases, we consider the diagrammatic construction for the
single photon and the two photon $S$-matrix elements by the connected $T$%
-matrix elements below. For the single photon case, the $S$-matrix element%
\begin{equation}
S_{p;k}=\delta _{kp}+iT_{p;k}\,,  \label{S1}
\end{equation}%
is defined by the $T$-matrix element for single photon, where $k$ and $p$
are the momenta of the incoming and outgoing photons. In this case, there
exist two kinds of disconnected diagrams (a) and (b) as shown in Fig.~\ref%
{fig3}.
\begin{figure}[tbp]
\includegraphics[bb=117 413 518 692, width=6 cm, clip]{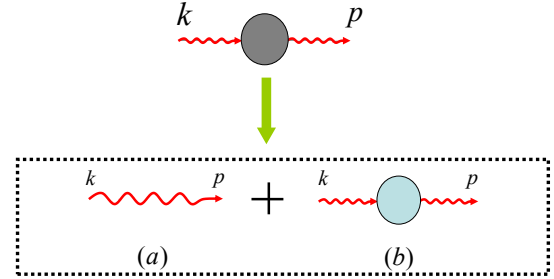}
\caption{(Color online) The diagrammatic constructions of the single photon $%
S$-matrix element: There exist two kinds of disconnected diagrams (a) and
(b). The red line denotes the free photon propagator. The gray shade circles
denote the $S$-matrix elements and the blue shade circles denote the single
photon $T$-matrix elements.}
\label{fig3}
\end{figure}

\begin{figure}[tbp]
\includegraphics[bb=65 144 567 695, width=7 cm, clip]{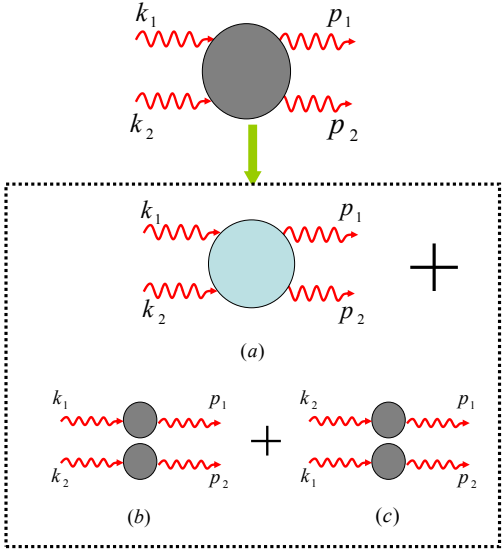}
\caption{(Color online) The diagrammatic constructions of the two photon $S$%
-matrix element: There exist three kinds of disconnected diagrams (a), (b)
and (c). The red line denotes the free photon propagator. The gray shade
circles denote the $S$-matrix elements and the blue shade circles denote the
two photon $T$-matrix elements.}
\label{fig4}
\end{figure}

For the two photon case, the $S$-matrix element%
\begin{equation}
S_{p_{1}p_{2};k_{1}k_{2}}=S_{p_{1}k_{1}}S_{p_{2}k_{2}}+S_{p_{2}k_{1}}S_{p_{1}k_{2}}+iT_{p_{1}p_{2};k_{1},k_{2}},
\label{S2}
\end{equation}%
is reduced by the $T$-matrix element of two photons, where $k_{r}$ and $%
p_{r} $ ($r=1,2$)\ are the momenta of the incoming and outgoing photons. In
this case, there exist three kinds of disconnected diagrams (a), (b) and (c)
as shown in Fig.~\ref{fig4}. Obviously, the multi-photon $S$-matrix is
totally determined by the connected $T$-matrix, so we only need to find the
photonic $T$-matrices.

Fortunately, the intrinsic relation%
\begin{equation}
iT_{2n}=\left. G_{2n}\prod\limits_{r=1}^{n}[2\pi
\,G_{0}^{-1}\!(k_{r})G_{0}^{-1}\!(p_{r})]\right\vert _{os}\,,  \label{T}
\end{equation}%
between $n$-photon $T$-matrix element%
\begin{equation}
T_{2n}=T_{p_{1},...,p_{n};k_{1},...,k_{n}}\,,
\end{equation}%
and photonic Green's function%
\begin{equation}
G_{2n}=G_{p_{1},...,p_{n};k_{1},...,k_{n}}\,,
\end{equation}%
is given by the LSZ reduction formula, where%
\begin{equation}
G_{0}(k_{r})=\frac{i}{\omega _{r}-\varepsilon _{k_{r}}+i0^{+}},
\end{equation}%
is the Green's function of free photon and $%
G_{p_{1},...,p_{n};k_{1},...,k_{n}}$ is the Fourier transformation of Eq.~(%
\ref{Green}). Here, the subscript $os$ denotes the on shell limit $\omega
\rightarrow \varepsilon _{k}$. Finally, the multi-photon $S$-matrix elements
are determined by the photonic Green's functions entirely. In this paper, we
use an elegant method, i.e., functional integrals, to establish the desired
relation~(\ref{T}). This method has been used to solve many quantum impurity
problem~\cite{Andrei1,Andrei2} in the condensed matter physics. Here, we
show how it works for the multi-photon transmission in this hybrid system.

\subsection{General formula for $S$-matrix in the T-type CRA}

We utilize the generating functional $Z$ to represent the full time ordering
Green's function as%
\begin{equation}
G_{p_{1},...,p_{n};k_{1},...,k_{n}}=\left. \frac{(-1)^{n}\,\delta ^{2n}\!\ln
Z\![\eta _{k},\eta _{k}^{\ast }]}{\delta \eta _{p_{1}}^{\ast }...\delta \eta
_{p_{n}}^{\ast }\delta \eta _{k_{1}}...\delta \eta _{k_{n}}}\right\vert
_{\eta _{k}=\eta _{k}^{\ast }=0}\,,  \label{G}
\end{equation}%
where the generating functional%
\begin{eqnarray}
Z[\eta ,\eta ^{\ast }] &=&\int D\![a_{k}a_{k}^{\dagger }]D\![f_{\sigma
}f_{\sigma }^{\dagger }]\,\delta \!(\sum_{\sigma }f_{\sigma }^{\dagger
}f_{\sigma }-1)  \notag \\
&&\exp \{i[S+\int dt\,\sum_{k}(\eta _{k}^{\ast }a_{k}+\eta
_{k}a_{k}^{\dagger })]\}\,,  \label{Z}
\end{eqnarray}%
is defined by the action $S=\int dt\,L$ and the Lagrangian%
\begin{equation}
L=if_{e}^{\dagger }\partial _{t}f_{e}+if_{g}^{\dagger }\partial
_{t}f_{g}+i\sum_{k}a_{k}^{\dagger }\partial _{t}a_{k}-H_{T}\,.
\end{equation}%
In Eq.~(\ref{Z}), we represent the two- level atom by the fermions $%
f_{\sigma }$ as%
\begin{eqnarray}
\left\vert e\right\rangle \left\langle e\right\vert &=&f_{e}^{\dagger
}f_{e}\,,  \notag \\
\left\vert e\right\rangle \left\langle g\right\vert &=&f_{e}^{\dagger
}f_{g}\,,
\end{eqnarray}%
with the constraint%
\begin{equation}
\sum_{\sigma =e,g}f_{\sigma }^{\dagger }f_{\sigma }=1\,.  \label{C}
\end{equation}%
This constraint arises from the fact that the physical space of the atom
spanned by $\left\vert e\right\rangle $ and $\left\vert g\right\rangle $ is
two dimensions, while the physical space of the fermions spanned by $%
\left\vert 0\right\rangle $, $f_{e}^{\dagger }\left\vert 0\right\rangle $, $%
f_{g}^{\dagger }\left\vert 0\right\rangle $ and $f_{e}^{\dagger
}f_{g}^{\dagger }\left\vert 0\right\rangle $ is four dimensions. By
integrating the photon field and the fermionic fields $f_{\sigma }$ in Eq.~(%
\ref{Z}), the generating functional is obtained as%
\begin{eqnarray}
\ln Z\![\eta ,\eta ^{\ast }] &=&Tr\ln M\![\xi ,\xi ^{\ast }]  \notag \\
&&-i\int d\omega dk\,\frac{\left\vert \eta _{k}\!(\omega )\right\vert ^{2}}{%
\omega -\varepsilon _{k}+i0^{+}}\,,  \label{GF}
\end{eqnarray}%
where the field variable is%
\begin{equation}
\xi \!(\omega )=V\int \frac{dk}{2\pi }\frac{\eta _{k}\!(\omega )}{\omega
-\varepsilon _{k}+i0^{+}}\,,
\end{equation}%
and the matrix%
\begin{equation}
M\![\xi ,\xi ^{\ast }]=\left(
\begin{array}{cc}
\lbrack \omega -\Omega +\Sigma \!(\omega )]\delta _{\omega \omega ^{\prime }}
& \xi \!(\omega -\omega ^{\prime }) \\
\xi ^{\dagger }\!(\omega ^{\prime }-\omega ) & (\omega -i0^{+})\delta
_{\omega \omega ^{\prime }}%
\end{array}%
\right) \,,
\end{equation}%
is defined by the self-energy%
\begin{equation}
\Sigma (\omega )={Re}\Sigma (\omega )+i\frac{\Gamma }{2}\rho (\omega )\,,
\end{equation}%
of the atom and $\Gamma =V^{2}$. Here, the real part of the self-energy is
determined by the principal-value integral as%
\begin{equation}
{Re}\Sigma (\omega )=\int \frac{dk}{2\pi }P\frac{\Gamma }{\varepsilon
_{k}-\omega }\,,
\end{equation}%
and the imaginary part is proportional to the density of state (DOS)%
\begin{equation}
\rho (\omega )=\sum_{i}\frac{1}{\left\vert D_{i}(\omega )\right\vert }\,,
\end{equation}%
where $D_{i}(\omega )=\left. \partial _{k}\varepsilon _{k}\right\vert
_{k=z_{i}}$ and $z_{i}$ is the real root of the equation $\varepsilon
_{z_{i}}=\omega $. For the waveguide the self-energy $\Sigma (\omega )$ is a
constant. For the CRA the self-energy depends on the frequency $\omega $ and
we have used Markov approximation to obtain Eq.~(\ref{GF}).

With the helps of Eqs.~(\ref{T}),~(\ref{G}) and~(\ref{GF}), we can achieve
the multi-photon $S$-matrix by considering the Green's function $G_{2n}$. In
the following, we use the LSZ approach to study the multi-photon transport
in the complex CRA.

\section{Bound states of the single photon transport in the T-type CRA}

In this section, we consider the single photon transport in the T-type CRA.
This problem has been considered in Ref.~\cite{Zhou}, but the photon bound
states have not been taken into account. There are two interesting physical
phenomena about the photon bound states. (a) Firstly, in the hybrid system
the single photon bound states depict the states that the single photon is
almost localized in the cavity containing the two-level atom. In this sense,
the photon bound states can be used as the quantum information storage of
single photon. (b) Secondly, as the basic element of single photon
transistor, the T-type CRA is usually coupled with other CRAs through the
two-level atom. These CRAs can be regarded as the scattering channels. It is
proved that~\cite{Xu} the incident photon in another CRA, whose energy is
resonance on the bound state energy in the T-type CRA, will be totally
reflected. This phenomenon can be considered as the one-dimension photonic
Feshbach resonance, which can be used to simulate the Feshbach resonance in
the atomic and condensed matter physics. The above reasons motivate us to
study the photon bound states in the T-type CRA by using the LSZ approach.

The T-type CRA is described by the Hamiltonian~(\ref{HT1}). Then Eqs.~(\ref%
{T}),~(\ref{G}) and~(\ref{GF}) immediately give the single photon $S$-matrix
element%
\begin{equation}
S_{p;k}=(1+r_{k})\delta _{kp}+r_{k}\delta _{-kp}\,,  \label{ST}
\end{equation}%
with the reflection amplitude%
\begin{equation}
r_{k}=\frac{-i\Gamma }{2J\sin k\,(\varepsilon _{k}-\Omega )+i\Gamma }\,.
\end{equation}%
Here, we have utilized DOS $\rho (\omega )=2/\sqrt{4J^{2}-(\omega -\omega
_{0})^{2}}$ and ${Re}\Sigma (\omega )=0$ for the scattered photon in the
T-type CRA. The result~(\ref{ST}) is the same as that obtained in Ref.~\cite%
{Zhou}.

\begin{figure}[tbp]
\includegraphics[bb=41 266 567 658, width=8 cm, clip]{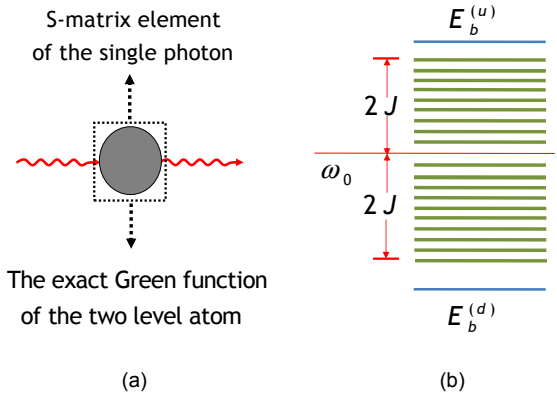}
\caption{(Color online) (a) The gray shade circle denotes the scattering
matrix element of the single photon, and it is also the Green function of
the two level atom. The poles of the atom Green function determine the
energies of the bound states. (b) The band structure of T-type CRA is shown
in the right panel.}
\label{fig5}
\end{figure}

\begin{figure}[tbp]
\includegraphics[bb=0 0 431 279, width=8 cm, clip]{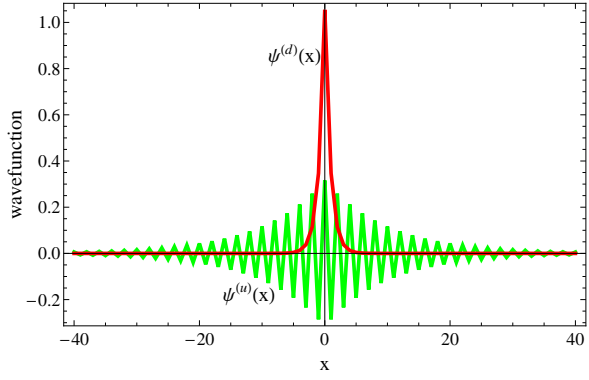}
\caption{(Color online) The wave-functions of two photonic bound state in
the discrete coordinates. The norms of both wave-functions exponentially
decay as $\left\vert x\right\vert $ increases.}
\label{fig6}
\end{figure}

In the following, we study the single photon bound states in this system by
considering the poles of $S$-matrix (see Fig.~\ref{fig5}a).
The poles of $S$-matrix are the roots of the equation%
\begin{equation}
(E_{b}-\Omega )+i\frac{\Gamma }{2}\Sigma \!(E_{b})=0\,.  \label{EB}
\end{equation}%
Because the energies of bound states are not inside the photonic energy band
of the CRA, i.e., $\left\vert E_{b}-\omega _{0}\right\vert >2J$, the
self-energy is%
\begin{equation}
\Sigma (E_{b})=-\frac{\Gamma \,sign\!(E_{b}-\omega _{0})}{\sqrt{%
(E_{b}-\omega _{0})^{2}-4J^{2}}}\,.
\end{equation}%
Then Eq.~(\ref{EB}) becomes%
\begin{equation}
E_{b}-\Omega -\frac{\Gamma \,sign\!(E_{b}-\omega _{0})}{\sqrt{(E_{b}-\omega
_{0})^{2}-4J^{2}}}=0\,.  \label{EB1}
\end{equation}%
The above equation possesses two solutions ($E_{b}^{(d)}$ and $E_{b}^{(u)}$)
for $E_{b}$: one solution $E_{b}^{(d)}$ is below the bottom of energy band,
the other solution $E_{b}^{(u)}$ is above the top of energy band. The
structure of the energy spectrum is shown in Fig~\ref{fig5}b. The two bound
states are%
\begin{equation}
\left\vert B_{\pm }\right\rangle =[\sum_{i}\psi _{\pm }(x_{i})a_{i}^{\dagger
}+\sigma ^{+}]\left\vert 0\right\rangle \,,
\end{equation}%
with the wave-functions~\cite{Xu} in the spatial representation%
\begin{equation}
\begin{array}{c}
\psi _{-}(x_{i})=\frac{(-)^{\left\vert x\right\vert }V}{\sqrt{%
(E_{b}^{(u)}-\omega _{0})^{2}-4J^{2}}}e^{\left\vert x\right\vert \ln \kappa
_{-}\!(E_{b}^{(u)})},\text{ if }E_{b}=E_{b}^{(u)} \\
\psi _{+}(x_{i})=\frac{V}{\sqrt{(E_{b}^{(d)}-\omega _{0})^{2}-4J^{2}}}%
e^{\left\vert x\right\vert \ln \kappa _{+}\!(E_{b}^{(d)})},\text{ if }%
E_{b}=E_{b}^{(d)}%
\end{array}%
\,,  \label{WF}
\end{equation}%
where%
\begin{equation}
\kappa _{\pm }\!(E)=-\sqrt{(\frac{E-\omega _{0}}{2J})^{2}-1}\pm \frac{\omega
_{0}-E}{2J}\,.
\end{equation}%
It is obviously that $\ln \kappa _{\pm }\!(E)<0$ for the both cases $%
E=E_{b}^{(u)}$ and $E=E_{b}^{(d)}$; the wave functions exponentially decay
as $\left\vert x\right\vert $ increases as shown in Eq.~(\ref{WF}), which
insure that the photon is indeed localized to form the bound states. The
wave-functions $\psi _{\pm }(x_{i})$ are illustrated in Fig.~\ref{fig6}.

By short summary for this section, we verify the existence of photon bound
states in the T-type CRA by considering the poles of the $S$-matrix element.
In the next section, using the T-type CRA to simulate the T-type waveguide,
we study the multi-photon transport in the T-type waveguide.

\section{Scattering matrix for photons in the T-type waveguide}

In this section, we focus on the photon transport in the T-type waveguide.
For the T-type waveguide, $S$-matrix element is firstly calculated out by
the LSZ reduction. Secondly, by using the obtained $S$-matrix, we give the
out-state for arbitrary incident state of photons. Finally, by analyzing the
photon out-state in the spatial representation, we can obtain the quantum
statistical properties of the scattered photons, such as the photon bunching
and anti-bunching.

As shown in Sec.~II, the waveguide is simulated by CRA in the limits $k\sim
\pm \pi /2$ and $\omega _{0}=\pi J$, in which the dispersion relation of the
photon is $\varepsilon _{k}=v_{g}\left\vert k\right\vert =\left\vert
k\right\vert $. Here, we let the group velocity $v_{g}=1$ for convenient.
Then the Hamiltonian%
\begin{eqnarray}
H_{T} &=&\Omega \left\vert e\right\rangle \!\left\langle e\right\vert
+\sum_{k}\left\vert k\right\vert a_{k}^{\dagger }a_{k}  \notag \\
&&+\frac{V}{\sqrt{L}}\sum_{k}(a_{k}^{\dagger }\sigma ^{-}+\text{\textrm{H.c.}%
})\,,
\end{eqnarray}%
describing the T-type waveguide becomes%
\begin{equation}
H_{T}^{(w)}=H_{T}^{e}+H_{T}^{o}\,,
\end{equation}%
with two parts%
\begin{equation}
H_{T}^{o}=\sum_{k>0}\varepsilon _{k}a_{k,o}^{\dagger }a_{k,o}\,,
\end{equation}%
and%
\begin{eqnarray}
H_{T}^{e} &=&\Omega \left\vert e\right\rangle \left\langle e\right\vert
+\sum_{k>0}ka_{k,e}^{\dagger }a_{k,e}  \notag \\
&&+\frac{\tilde{V}}{\sqrt{L}}\sum_{k>0}(a_{k,e}^{\dagger }\sigma ^{-}+\text{%
\textrm{H.c.}})\,,  \label{HTe}
\end{eqnarray}%
where the operators%
\begin{eqnarray}
a_{k,e} &=&\frac{1}{\sqrt{2}}(a_{k}+a_{-k})\,,  \notag \\
a_{k,o} &=&\frac{1}{\sqrt{2}}(a_{k}-a_{-k})\,,  \label{a}
\end{eqnarray}%
describe annihilations of the $e$-photon and $o$-photon. Here, $e$-photon
depicts the photon with even parity in the momentum space, i.e., $%
a_{-k,e}=a_{k,e}$, and $o$-photon depicts the photon with odd parity in the
momentum space, i.e., $a_{-k,o}=-a_{k,o}$. In Eq.~(\ref{HTe}), the effective
coupling constant is $\tilde{V}=\sqrt{2}V$. For the $o$-photon, the $%
H_{T}^{o}$ is diagonalized in the bases \{$a_{k,o}^{\dagger }\left\vert
0\right\rangle $\}, so we only need to find out the $S$-matrix for the the
scattered $e$-photon. In the following, we consider the $S$-matrix for the $%
e $-photon according to the approach.

\subsection{Single photon scattering}

For the single $e$-photon case, Eqs.~(\ref{G}) and~(\ref{GF}) give the
single photon Green's function%
\begin{equation}
G\!(p;k)=[G_{0}\!(k)-\frac{i}{2\pi }\frac{\Gamma _{T}}{\omega _{k}-\alpha }%
G_{0}^{2}\!(k)]\delta _{pk}\,,  \label{G1}
\end{equation}%
where $\alpha =\Omega -i\Gamma _{T}/2$ and $\Gamma _{T}=\tilde{V}^{2}$.
Here, we used ${Re}\Sigma (\omega )=0$ and $\rho (\omega )=1$. Together with
Eq.~(\ref{T}), Eq.~(\ref{G1}) gives the single $e$-photon $T$-matrix%
\begin{equation}
iT_{p;k}=\delta _{pk}\frac{-i\Gamma _{T}}{k-\alpha }\,.
\end{equation}%
Next, we achieve the single $e$-photon $S$-matrix element $%
S_{p;k}=t_{k}\delta _{pk}$ by considering Eq.~(\ref{S1}), where the
transmission coefficient is%
\begin{equation}
t_{k}=\frac{k-\alpha ^{\ast }}{k-\alpha }\,.  \label{tk}
\end{equation}%
This result~(\ref{tk}) accords with that of Refs.~\cite{FanPRL,FanPRA} based
on the Lippmann-Schwinger formalism.

\subsection{Two photon scattering}

For the two $e$-photon case, Eqs.~(\ref{G}) and~(\ref{GF}) give the two $e$%
-photon Green's function as follows:%
\begin{eqnarray}
G_{p_{1},p_{2};k_{1},k_{2}} &=&i\frac{2\Gamma _{T}^{2}}{(2\pi )^{3}}\frac{%
G_{0}\!(k_{1})G_{0}\!(k_{2})G_{0}\!(p_{1})G_{0}\!(p_{2})}{(\omega
_{2}^{\prime }-\alpha )(\omega _{1}^{\prime }-\alpha )}  \notag \\
&&\times \frac{(\omega _{1}+\omega _{2}-2\alpha )\delta _{\omega _{1}+\omega
_{2},\omega _{1}^{\prime }+\omega _{2}^{\prime }}}{(\omega _{1}-\alpha
)(\omega _{2}-\alpha )}\,.
\end{eqnarray}%
By taking the photon frequency $\omega _{i}$ and $\omega _{i}^{\prime }$ on
shell, we obtain the $T$-matrix%
\begin{eqnarray}
iT_{p_{1},p_{2};k_{1},k_{2}} &=&i\frac{\Gamma _{T}^{2}}{\pi }\frac{%
(k_{1}+k_{2}-2\alpha )}{(p_{2}-\alpha )(k_{1}-\alpha )}  \notag \\
&&\times \frac{\delta _{k_{1}+k_{2},p_{1}+p_{2}}}{(p_{1}-\alpha
)(k_{2}-\alpha )}\,.
\end{eqnarray}%
From the above equation, the two $e$-photon $S$-matrix element%
\begin{eqnarray}
S_{p_{1}p_{2};k_{1}k_{2}} &=&iT_{p_{1}p_{2};k_{1},k_{2}}  \notag \\
&&+t_{k_{1}}t_{k_{2}}(\delta _{p_{1}k_{1}}\delta _{p_{2}k_{2}}+\delta
_{p_{2}k_{1}}\delta _{p_{1}k_{2}})\,.
\end{eqnarray}%
follows Eq.~(\ref{S2}) immediately. If two incident photons are prepared in
the state $\left\vert k_{1},k_{2}\right\rangle $, the wave function~\cite%
{FanPRL,FanPRA}%
\begin{eqnarray}
\left\langle x_{c},x\left\vert out\right\rangle \right. &=&\frac{1}{2}%
\sum_{p_{1}p_{2}}S_{p_{1}p_{2};k_{1}k_{2}}\left\langle x_{c},x\left\vert
p_{1},p_{2}\right\rangle \right.  \notag \\
&=&e^{iEx_{c}}\frac{1}{2\pi }[t_{k_{1}}t_{k_{2}}\cos (\Delta _{k}x)  \notag
\\
&&-\frac{4\Gamma _{T}^{2}e^{i(E-2\Omega +i\Gamma _{T})\left\vert
x\right\vert /2}}{4\Delta _{k}^{2}-(E-2\Omega +i\Gamma _{T})^{2}}]\,,
\label{outT}
\end{eqnarray}%
of two outgoing photons in the spatial representation is obtained in terms
of the two $e$-photon center of mass coordinate $x_{c}=(x_{1}+x_{2})/2$ and
two photon relative coordinate $x=x_{1}-x_{2}$. Here, the total momentum
(energy) is $E=k_{1}+k_{2}$ and the relative momentum is $\Delta
_{k}=(k_{1}-k_{2})/2$. When the photon momenta $k_{1}$ and $k_{2}$ both
satisfy the resonance condition $k_{1}=k_{2}=\Omega $, the envelop
wave-function~(\ref{outT}) exponentially decays as the relative coordinate $%
x $ increases. This reflects that the outgoing two photons attract with each
other effectively and form a two photon bound state. If $E-2\Omega $ is kept
to zero, the wave-function at $x=0$ decreases as $\left\vert \Delta
_{k}\right\vert $ increases, which implies the photons repulse against each
other effectively through interacting with the two- level atom. The above
results about two photon scattering accord with the results reported in
Refs.~\cite{FanPRL,FanPRA}.
\begin{figure}[tbp]
\includegraphics[bb=91 120 519 666, width=6 cm, clip]{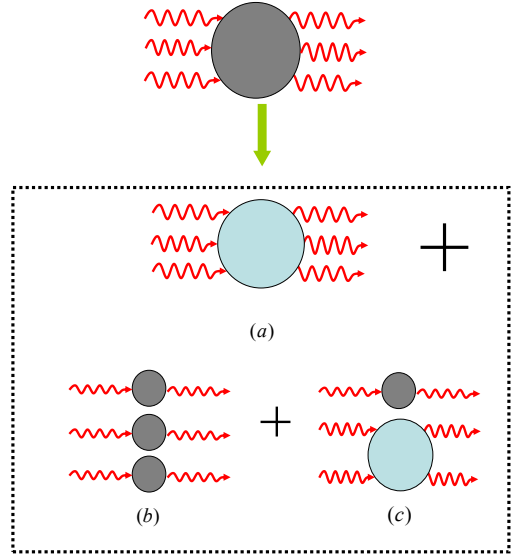}
\caption{(Color online) Feynman diagrams for three photon $S$ matrix: There
exist three kinds of disconnected diagrams (a), (b) and (c).}
\label{fig7}
\end{figure}

\subsection{Three photon scattering}

For the three photon case, Eqs. (\ref{T}) and (\ref{G}) give the connected $%
T $-matrix as%
\begin{equation}
iT_{p_{1}p_{2}p_{3};k_{1}k_{2}k_{3}}=i\frac{\Gamma _{T}^{3}}{3(2\pi )^{2}}%
\delta \lbrack \sum_{i}(k_{i}-p_{i})]\sum_{a=1,2,3}F_{k_{i},p_{i}}^{(a)}\,,
\end{equation}%
where the functions $F_{\omega _{i},\omega _{i}^{\prime }}^{(a)}$ are
defined as%
\begin{eqnarray}
F_{\omega _{i},\omega _{i}^{\prime }}^{(1)} &=&\sum_{PQ}\frac{%
\prod_{i}\delta (\omega _{i}-k_{P_{i}})}{(\omega _{1}^{\prime }-\omega
_{1})(\omega _{3}^{\prime }-\omega _{3})(\omega _{1}-\alpha )}  \notag \\
&&\frac{\prod_{i}\delta (\omega _{i}^{\prime }-p_{Q_{i}})}{(\omega
_{3}^{\prime }-\alpha )(\omega _{1}+\omega _{2}-\omega _{1}^{\prime }-\alpha
)}\,,
\end{eqnarray}%
\begin{eqnarray}
F_{\omega _{i},\omega _{i}^{\prime }}^{(2)} &=&\sum_{PQ}\frac{%
\prod_{i}\delta (\omega _{i}-k_{P_{i}})}{(\omega _{2}^{\prime }-\omega
_{2})(\omega _{3}^{\prime }-\omega _{3})(\omega _{2}^{\prime }-\alpha )}
\notag \\
&&\frac{\prod_{i}\delta (\omega _{i}^{\prime }-p_{Q_{i}})}{(\omega
_{3}-\alpha )(\omega _{2}^{\prime }+\omega _{1}^{\prime }-\omega _{2}-\alpha
)}\,,
\end{eqnarray}%
and%
\begin{eqnarray}
F_{\omega _{i},\omega _{i}^{\prime }}^{(3)} &=&\sum_{PQ}\frac{%
\prod_{i}\delta (\omega _{i}-k_{P_{i}})}{(\omega _{2}^{\prime }-\omega
_{2})(\omega _{1}^{\prime }-\omega _{1})(\omega _{1}^{\prime }-\alpha )}
\notag \\
&&\frac{\prod_{i}\delta (\omega _{i}^{\prime }-p_{Q_{i}})}{(\omega
_{2}-\alpha )(\omega _{2}+\omega _{3}-\omega _{2}^{\prime }-\alpha )}\,.
\end{eqnarray}%
Here, $P=(P_{1},P_{2},P_{3})$ and $Q=(Q_{1},Q_{2},Q_{3})$ are two different
permutations of ($1,2,3$), and $i=1,2,3$. The three photon $S$-matrix element%
\begin{eqnarray}
S_{p_{1},p_{2},p_{3};k_{1},k_{2},k_{3}} &=&i\sum_{i,j=1,2,3}\sum_{\gamma
,\delta \neq i}\sum_{\lambda ,\beta \neq j}S_{p_{j};k_{i}}T_{p_{\lambda
},p_{\beta };k_{\gamma },k_{\delta }}  \notag \\
&&+\sum_{PQ}\prod_{i=1,2,3}S_{p_{Q_{i}};k_{P_{i}}}  \notag \\
&&+iT_{p_{1},p_{2},p_{3};k_{1},k_{2},k_{3}}\,,
\end{eqnarray}%
is constructed by summing up the contributions from all disconnected Feynman
diagrams as shown in Fig.~\ref{fig7}. Here, $\gamma ,\delta ,\lambda ,\beta
\in \{1,2,3\}$. The similar discussions can be applicable to deal with the $%
N $-photon scattering process.
\begin{figure}[tbp]
\includegraphics[bb=8 337 590 701, width=8 cm, clip]{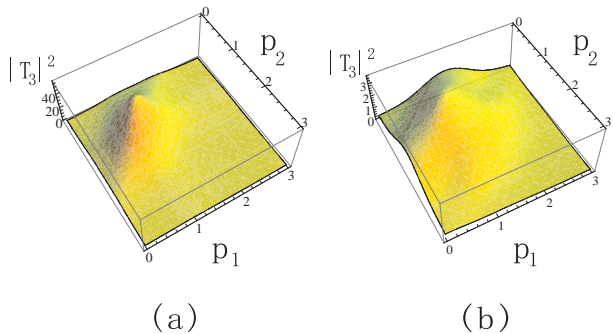}
\caption{(Color online) Three-photon background fluorescence for the total
energy of incident photons $E=k_{1}+k_{2}+k_{3}=3$ and $\Gamma _{T}=1$,
where the energy level spacing $\Omega $ is taken as units: (a) the three
photon are both on resonance with the atom, i.e., $k_{1}=k_{2}=k_{3}=\Omega $%
; (b) the energies of the three photons are $k_{1}=0.5$, $k_{2}=0.3$, and $%
k_{3}=2.2$, respectively.}
\label{fig8}
\end{figure}

Next, we consider the physical meaning of $S$-matrix. To this end, we first
analysis the $T$-matrices, which are $\left\vert T_{2}\right\vert
^{2}=\left\vert T_{p_{1}p_{2};k_{1},k_{2}}\right\vert ^{2}$ and $\left\vert
T_{3}\right\vert ^{2}=\left\vert
T_{p_{1},p_{2},p_{3};k_{1},k_{2},k_{3}}\right\vert ^{2}$. Here, $\left\vert
T_{2}\right\vert ^{2}$ describes the two-photon background fluorescence,
which is explicitly discussed in Ref.~\cite{FanPRL,FanPRA}, and $\left\vert
T_{3}\right\vert ^{2}$ depicts the three-photon background fluorescence,
which is shown in Fig.~\ref{fig8}. It is shown that when the three photons
are all on resonance with the atom, the three-photon background fluorescence
describing $\left\vert T_{3}\right\vert ^{2}$ is enhanced largely.

\begin{figure}[tbp]
\includegraphics[bb=35 26 539 807, width=8 cm, clip]{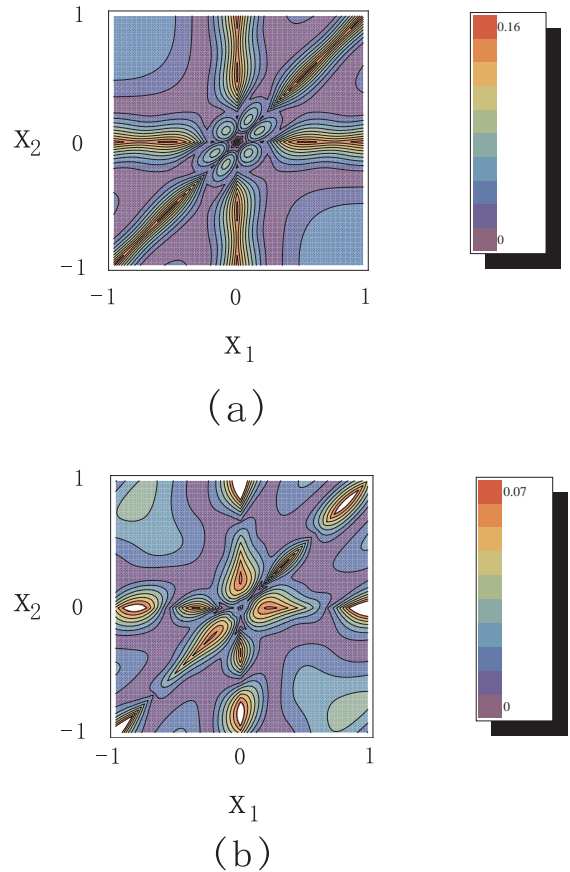}
\caption{(Color online) The probability distribution of three photon with
one photon at origin, where the total energy of incident photons $%
E=k_{1}+k_{2}+k_{3}=3$, the energy level spacing $\Omega $ is taken as
units, and $\Gamma _{T}=1$: (a) the three photon are both on resonance with
the atom, i.e., $k_{1}=k_{2}=k_{3}=\Omega $; (b) the energies of the three
photons are $k_{1}=0.5$, $k_{2}=1.5$, and $k_{3}=1$, respectively.}
\label{fig9}
\end{figure}

The out-going state of three photons%
\begin{equation}
\left\vert out\right\rangle =\sum_{p_{1}\leq p_{2}\leq
p_{3}}S_{p_{1},p_{2},p_{3};k_{1},k_{2},k_{3}}\left\vert
p_{1},p_{2},p_{3}\right\rangle \,,
\end{equation}%
is determined by the three photon $S$-matrix, and its spatial representation
the wave-function reads as%
\begin{eqnarray}
&&\left\langle x_{1},x_{2},x_{3}\left\vert out\right\rangle \right.  \notag
\\
&=&\sum_{p_{1}p_{2}p_{3}}\frac{S_{p_{1},p_{2},p_{3};k_{1},k_{2},k_{3}}}{%
6(2\pi )^{3/2}}e^{i(p_{1}x_{1}+p_{2}x_{2}+p_{3}x_{3})}\,.
\end{eqnarray}%
The contour maps of the probability distributions $\left\vert \left\langle
x_{1},x_{2},x_{3}\left\vert out\right\rangle \right. \right\vert ^{2}$ are
numerically shown in Fig.~\ref{fig9}. It is illustrated in Fig.~\ref{fig9}a
that when the three photon are all on resonance with the atom, the scattered
photons prefer the two photon bound state rather than the three photon bound
state. That is, if two photons form the bound state, it is difficult to form
three photon bound state, namely, the two bounded photons repulse another
one effectively. When the three photon are not on resonance, the probability
distribution $\left\vert \left\langle x_{1},x_{2},x_{3}\left\vert
out\right\rangle \right. \right\vert ^{2}$ is shown in Fig.~\ref{fig9}b. It
is illustrated in Fig.~\ref{fig9}b that it is also difficult to realize the
three photon bound state. If the position of one photon is given, such as $%
x_{3}=0$, other two photons do not always attract or repulse each other, but
attract each other at some points and repulse each other at other points,
which are determined by the distance between the two photon and another
photon localized at $x_{3}=0$.
It is follows the above discussion that, in the conceptual setup of the
photon transistor, the two- level atom controls the coherent transport
behaviors of single photon, such as the transmission and reflection. In the
multi-photon transport, the atom can induce the effective interaction of
photons. We can control the effective interaction by adjusting the energy
level spacing of the atom. Therefore, the coherent manipulations of TLS can
result in a transitions from the repulsive case to attraction of effective
photon interactions.

In addition, we point that the recent references.~\cite{FanPRL,FanPRA}
obtained the same results for two photons transport, but our works are
different from them: (1) Refs.~\cite{FanPRL,FanPRA} only give the two photon
eigenstates by Bethe-ansatz method, but we find a general method~\cite%
{Andrei-Bethe, Andrei-Bethe-2}, i.e., the scattering Bethe ansatz technique
(see Appendix), to derive the multi-photon eigenstates; (2) though we can
obtain the multi-photon eigenstates by the subtle scattering Bethe-ansatz
method, we still need a lot of complicated calculations to achieve the $S$%
-matrix by using the Lippmann-Schwinger scattering theory. However, the LSZ
approach can be generalized to study the multi-photon scattering in the
waveguide, such as the three photon transport in the waveguide; (3) Except
these, the LSZ approach is also used to deal with the multi-photon
scattering in the more complex CRA, such as the H-type CRA. In the next
section, by using the H-type CRA to simulate the H-type waveguide, we
investigate the multi-photon transport in the H-type waveguide.

\section{Two photon scattering process in the H-type waveguide}

In this section, we study the two photon scattering process in the H-type
waveguide.
\begin{figure}[tbp]
\includegraphics[bb=91 210 519 666, width=6 cm, clip]{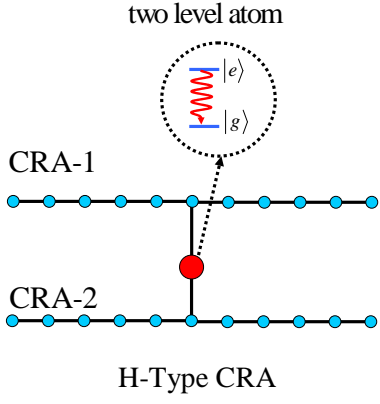}
\caption{(Color online) The schematic for the H-type CRA is shown in this
figure. The red circle denotes the two level atom. The blue dots denote the
coupled resonators.}
\label{fig10}
\end{figure}
The conventional H-type waveguide is simulated by the H-type CRA (Fig.~\ref%
{fig10}) in the high energy limits. The model Hamiltonian%
\begin{eqnarray}
H_{H} &=&\Omega \left\vert e\right\rangle \left\langle e\right\vert
+\sum_{i,s=1,2}\delta _{i0}(V_{s}a_{i,s}^{\dagger }\sigma ^{-}+h.c.)
\label{HH1} \\
&&+\sum_{i,s=1,2}[\omega _{0}^{(s)}a_{i,s}^{\dagger
}a_{i,s}-J_{s}(a_{i,s}^{\dagger }a_{i+1,s}+h.c.)]\,,  \notag
\end{eqnarray}%
of the H-type CRA is defined by the hopping constant $J_{s}$ and the
creation operator $a_{i,s}^{\dagger }$ of the $i$-th single mode cavity with
frequency $\omega _{0}^{(s)}$ in the CRA-$s$, where $V_{s}$ is the
hybridization constant of localized atom-photon in the $0$-th site of the
CRA-$s$. Here, $s$ denotes the CRA-1 or the CRA-2 as shown in Fig.~\ref%
{fig10}. In the $k$-space, the Hamiltonian (\ref{HH1}) becomes%
\begin{eqnarray}
H_{H} &=&\Omega \left\vert e\right\rangle \left\langle e\right\vert
+\sum_{k,s=1,2}\varepsilon _{k}^{(s)}a_{k,s}^{\dagger }a_{k,s}  \notag \\
&&+\frac{1}{\sqrt{L}}\sum_{k,s=1,2}(V_{s}a_{k,s}^{\dagger }\sigma
^{-}+h.c.)\,,  \label{Hk}
\end{eqnarray}%
where the dispersion relation of photon is $\varepsilon _{k}^{(s)}=\omega
_{0}^{(s)}-2J_{s}\cos k$.

In the high energy limits $k\longrightarrow \pm \pi /2$ and $\omega
_{0}^{(s)}=\pi J_{s}$, the dispersion relation of the photon is $\varepsilon
_{k}^{(s)}\sim v_{s}\left\vert k\right\vert $ with the group velocity $%
v_{s}=2J_{s}$. In the case, the Hamiltonian (\ref{Hk}) describes the photon
transport in the H-type waveguide. For convenient, we use the operators%
\begin{eqnarray}
a_{k,e}^{(s)} &=&\frac{1}{\sqrt{2}}(a_{k,s}+a_{-k,s})\,,  \notag \\
a_{k,o}^{(s)} &=&\frac{1}{\sqrt{2}}(a_{k,s}-a_{-k,s})\,,
\end{eqnarray}%
to rewrite the Hamiltonian (\ref{Hk}) as $H_{H}=H_{H}^{(e)}+H_{H}^{(o)}$,
where
\begin{equation}
H_{H}^{(o)}=\sum_{k>0,s=1,2}\varepsilon _{k}^{(s)}a_{k,o}^{(s)\dagger
}a_{k,o}^{(s)}\,,
\end{equation}%
for the $o$-photon and%
\begin{eqnarray}
H_{H}^{(e)} &=&\Omega \left\vert e\right\rangle \left\langle e\right\vert
+\sum_{k>0,s=1,2}\varepsilon _{k}^{(s)}a_{k,e}^{(s)\dagger }a_{k,e}^{(s)}
\notag \\
&&+\frac{1}{\sqrt{L}}\sum_{k>0,s=1,2}(\bar{V}_{s}a_{k,e}^{(s)\dagger }\sigma
^{-}+h.c.)\,,
\end{eqnarray}%
for the $e$-photon with the effective coupling $\bar{V}_{s}=\sqrt{2}V_{s}$.
Because Hamiltonian $H_{H}^{(o)}$ is diagonalized in the bases \{$%
a_{k,o}^{(s)\dagger }\left\vert 0\right\rangle $\}, so we only need to find
the $S$-matrix for the $e$-photon.

\subsection{The LSZ reduction for $e$-photon scattering in the H-type
waveguide}

For calculating the multi-photon $S$-matrix, we only consider the Green's
function and the $S$-matrix for the $e$-photon in the H-type waveguide. In
this case, the $2n$-point photonic Green's function reads%
\begin{eqnarray}
&&G_{p_{1},...,p_{n};k_{1},...,k_{n}}^{j_{1},...,j_{n};i_{1},...,i_{n}}(t_{1}^{\prime },...t_{n}^{\prime };t_{1},...t_{n})
\label{Green2} \\
&=&\left\langle Ta_{p_{1},e}^{(j_{1})}(t_{1}^{\prime
})...a_{p_{n},e}^{(j_{n})}(t_{n}^{\prime })a_{k_{1},e}^{(i_{1})\dagger
}(t_{1})...a_{k_{n},e}^{(i_{n})\dagger }(t_{n})\right\rangle _{H}\,.  \notag
\end{eqnarray}%
The $S$-matrix element is the overlap%
\begin{equation}
_{out}\left\langle f\left\vert i\right\rangle _{in}\right. =\left.
_{in}\left\langle f\right\vert S\left\vert i\right\rangle _{in}\right. \,,
\end{equation}%
of incoming wave $\left\vert i\right\rangle
_{in}=a_{k_{1},e}^{(i_{1})\dagger }...a_{k_{n},e}^{(i_{n})\dagger
}\left\vert 0\right\rangle $ and outgoing wave state $\left\vert
f\right\rangle _{out}$. As shown in the Sec. III, the basic part of the $S$%
-matrix is the $T$-matrix. The relation%
\begin{equation}
iT_{2n}^{(H)}=\left. G_{2n}^{(H)}\prod\limits_{r=1}^{n}[2\pi
G_{0i_{r}}^{-1}(k_{r})G_{0j_{r}}^{-1}(p_{r})]\right\vert _{os}\,,  \label{TH}
\end{equation}%
between the $2n$-point $T$-matrix element%
\begin{equation}
T_{2n}^{(H)}=T_{p_{1},...,p_{n};k_{1},...,k_{n}}^{j_{1},...,j_{n};i_{1},...,i_{n}}\,,
\end{equation}%
and the $2n$-point Green's function%
\begin{equation}
G_{2n}^{(H)}=G_{p_{1},...,p_{n};k_{1},...,k_{n}}^{j_{1},...,j_{n};i_{1},...,i_{n}}\,,
\end{equation}%
is determined by the LSZ reduction approach, where
\begin{equation}
G_{0i_{r}}(k_{r})=\frac{i}{\omega _{r}-\varepsilon _{k_{r}}^{(i_{r})}+i0^{+}}%
\,,
\end{equation}%
is the Green's function of the free photon and $%
G_{p_{1},...,p_{n};k_{1},...,k_{n}}^{j_{1},...,j_{n};i_{1},...,i_{n}}$ is
the Fourier transformation of Eq. (\ref{Green2}). Here, the subscript $os$
denotes the on shell limits $\omega \rightarrow \varepsilon _{k}^{(i)}$.
Then, the $2n$-point Green's function%
\begin{equation}
G_{p_{1},...,p_{n};k_{1},...,k_{n}}^{j_{1},...,j_{n};i_{1},...,i_{n}}=\left.
\frac{(-1)^{n}\delta ^{2n}\ln Z[\eta _{k}^{s},\eta _{k}^{s\ast }]}{\delta
\eta _{p_{1}}^{j_{1}\ast }...\delta \eta _{p_{n}}^{j_{n}\ast }\delta \eta
_{k_{1}}^{i_{1}}...\delta \eta _{k_{n}}^{i_{n}}}\right\vert _{\eta
_{k}^{s}=\eta _{k}^{s\ast }=0}\,,  \label{GH}
\end{equation}%
is obtained by the generating functional%
\begin{eqnarray}
\ln Z[\eta _{k}^{s},\eta _{k}^{s\ast }] &=&Tr\ln M[\xi ,\xi ^{\ast }]  \notag
\\
&&-i\sum_{s}\int d\omega dk\frac{\left\vert \eta _{k}^{s}(\omega
)\right\vert ^{2}}{\omega -\varepsilon _{k}^{(s)}+i0^{+}}\,,  \label{ZH}
\end{eqnarray}%
where we have used the matrix%
\begin{equation}
M[\xi ,\xi ^{\ast }]=\left(
\begin{array}{cc}
\lbrack \omega -\Omega +i\Gamma _{e}/2]\delta _{\omega \omega ^{\prime }} &
\xi (\omega -\omega ^{\prime }) \\
\xi ^{\dagger }(\omega ^{\prime }-\omega ) & (\omega -i0^{+})\delta _{\omega
\omega ^{\prime }}%
\end{array}%
\right) \,,
\end{equation}%
and the field variable%
\begin{equation}
\xi (\omega )=\sum_{s}\int \frac{dk}{2\pi }\frac{\bar{V}_{s}\eta
_{k}^{s}(\omega )}{\omega -\varepsilon _{k}^{(s)}+i0^{+}}\,.
\end{equation}%
Here, the atom decay rate is $\Gamma _{e}=\sum_{s}\bar{V}_{s}^{2}/v_{s}$.
Finally, together with Eqs. (\ref{GH}) and (\ref{ZH}), Eq. (\ref{TH}) gives
the basic element of the $S$-matrix, i.e., $T$-matrix. Using the $T$-matrix
elements, we can find the all $S$-matrix elements. For convenient, we let $%
v_{1}=v_{2}=1$ below.

\subsection{Single and two $e$-photon $S$-matrices}

We consider the single and the two $e$-photon $S$-matrices in this
subsection. The diagrammatic analysis shows that the single $e$-photon $S$%
-matrix element%
\begin{equation}
S_{p;k}^{j;i}=\delta _{kp}\delta _{ij}+iT_{p;k}^{j;i}\,,  \label{SH1}
\end{equation}%
consists of the $T$-matrix element%
\begin{equation}
iT_{p;k}^{j;i}=\left. G_{p;k}^{j;i}[2\pi
G_{0j}^{-1}(p)G_{0i}^{-1}(k)]\right\vert _{os}\,,
\end{equation}%
where $k$ and $p$ are the momenta of the incoming photon in the waveguide-$i$
and the outgoing photon in the waveguide-$j$ respectively. Together with Eq.
(\ref{GH}) and Eq. (\ref{ZH}), we obtain the single $e$-photon $T$-matrix
element as%
\begin{equation}
iT_{p;k}^{j;i}=\frac{-i\bar{V}_{i}\bar{V}_{j}}{k-\Omega +i\frac{1}{2}\Gamma
_{e}}\delta _{kp}\,.
\end{equation}%
Next, we consider the special case: the incident $e$-photon is prepared in
the waveguide-1. Then, the $S$-matrix elements are $%
S_{p;k}^{1;1}=t_{k}^{(11)}\delta _{kp}$ and $S_{p;k}^{2;1}=t_{k}^{(21)}%
\delta _{kp}$, for%
\begin{equation}
t_{k}^{(11)}=\frac{k-\Omega +i\frac{1}{2}(\bar{V}_{2}^{2}-\bar{V}_{1}^{2})}{%
k-\Omega +i\frac{1}{2}(\bar{V}_{2}^{2}+\bar{V}_{1}^{2})}\,,
\end{equation}%
and%
\begin{equation}
t_{k}^{(21)}=\frac{-i\bar{V}_{1}\bar{V}_{2}}{k-\Omega +i\frac{1}{2}\Gamma
_{e}}\,.
\end{equation}%
We can verify the unitarity of the $S$-matrix as%
\begin{equation}
\left\vert t_{k}^{(11)}\right\vert ^{2}+\left\vert t_{k}^{(21)}\right\vert
^{2}=1\,\text{.}
\end{equation}

For the two photon case, the diagrammatic analysis and the LSZ reduction
approach give the two $e$-photon $S$-matrix element%
\begin{equation}
S_{p_{1}p_{2};k_{1}k_{2}}^{j_{1},j_{2};i_{1},i_{2}}=S_{p_{1}k_{1}}^{j_{1};i_{1}}S_{p_{2}k_{2}}^{j_{2};i_{2}}+S_{p_{2}k_{1}}^{j_{2};i_{1}}S_{p_{1}k_{2}}^{j_{1};i_{2}}+iT_{p_{1}p_{2};k_{1},k_{2}}^{j_{1},j_{2};i_{1},i_{2}}\,,
\label{SH2}
\end{equation}%
with the two photon $T$-matrix element%
\begin{equation}
iT_{p_{1},p_{2};k_{1},k_{2}}^{j_{1},j_{2};i_{1},i_{2}}=\left.
G_{p_{1},p_{2};k_{1},k_{2}}^{j_{1},j_{2};i_{1},i_{2}}\prod%
\limits_{r=1}^{2}[2\pi
G_{0i_{r}}^{-1}(k_{r})G_{0j_{r}}^{-1}(p_{r})]\right\vert _{os}\,,
\end{equation}%
where $k_{r}$ and $p_{r}$ are the momenta of the incoming photons in the
waveguide-$i_{r}$ and the outgoing photons in the waveguide-$j_{r}$
respectively. Together with Eqs. (\ref{GH}) and (\ref{ZH}), the two $e$%
-photon $T$-matrix element becomes%
\begin{eqnarray}
iT_{p_{1}p_{2};k_{1},k_{2}}^{j_{1},j_{2};i_{1},i_{2}} &=&\frac{i\bar{V}%
_{i_{1}}\bar{V}_{i_{2}}\bar{V}_{j_{1}}\bar{V}_{j_{2}}(k_{1}+k_{2}-2\alpha
_{H})}{\pi (p_{2}-\alpha _{H})(k_{1}-\alpha _{H})}  \notag \\
&&\times \frac{\delta _{k_{1}+k_{2},p_{1}+p_{2}}}{(p_{1}-\alpha
_{H})(k_{2}-\alpha _{H})}\,,  \label{TH2}
\end{eqnarray}%
with $\alpha _{H}=\Omega -i\Gamma _{e}/2$.

Finally, we consider the special case: the two incident $e$-photons are
prepared in the waveguide-1 and the waveguide-2 respectively, i.e., the
incident state $\left\vert in\right\rangle $ is $\left\vert in\right\rangle
=a_{k_{1},e}^{(1)\dagger }a_{k_{2},e}^{(2)\dagger }\left\vert 0\right\rangle
$. In this case, Eqs. (\ref{SH2}) and (\ref{TH2}) give the $S$-matrix
elements as%
\begin{eqnarray}
S_{p_{1}p_{2};k_{1}k_{2}}^{1,1;1,2}
&=&t_{k_{1}}^{(11)}t_{k_{2}}^{(21)}(\delta _{k_{1}p_{1}}\delta
_{k_{2}p_{2}}+\delta _{k_{1}p_{2}}\delta _{k_{2}p_{1}})  \notag \\
&&+\frac{i\bar{V}_{2}\bar{V}_{1}^{3}}{\pi }\frac{(k_{1}+k_{2}-2\alpha _{H})}{%
(p_{2}-\alpha _{H})(k_{1}-\alpha _{H})}  \notag \\
&&\times \frac{\delta _{k_{1}+k_{2},p_{1}+p_{2}}}{(p_{1}-\alpha
_{H})(k_{2}-\alpha _{H})}\,,  \label{S1112}
\end{eqnarray}%
\begin{eqnarray}
S_{p_{1}p_{2};k_{1}k_{2}}^{1,2;1,2}
&=&t_{k_{1}}^{(11)}t_{k_{2}}^{(22)}\delta _{k_{1}p_{1}}\delta
_{k_{2}p_{2}}+t_{k_{1}}^{(21)}t_{k_{2}}^{(21)}\delta _{k_{1}p_{2}}\delta
_{k_{2}p_{1}}  \notag \\
&&+\frac{i\bar{V}_{1}^{2}\bar{V}_{2}^{2}}{\pi }\frac{(k_{1}+k_{2}-2\alpha
_{H})}{(p_{2}-\alpha _{H})(k_{1}-\alpha _{H})}  \notag \\
&&\times \frac{\delta _{k_{1}+k_{2},p_{1}+p_{2}}}{(p_{1}-\alpha
_{H})(k_{2}-\alpha _{H})}\,,  \label{S1212}
\end{eqnarray}%
and%
\begin{eqnarray}
S_{p_{1}p_{2};k_{1}k_{2}}^{2,2;1,2}
&=&t_{k_{1}}^{(21)}t_{k_{2}}^{(22)}(\delta _{k_{1}p_{1}}\delta
_{k_{2}p_{2}}+\delta _{k_{1}p_{2}}\delta _{k_{2}p_{1}})  \notag \\
&&+\frac{i\bar{V}_{1}\bar{V}_{2}^{3}}{\pi }\frac{(k_{1}+k_{2}-2\alpha _{H})}{%
(p_{2}-\alpha _{H})(k_{1}-\alpha _{H})}  \notag \\
&&\times \frac{\delta _{k_{1}+k_{2},p_{1}+p_{2}}}{(p_{1}-\alpha
_{H})(k_{2}-\alpha _{H})}\,.  \label{S2212}
\end{eqnarray}%
Here, $t_{k}^{(22)}$ is defined by%
\begin{equation}
t_{k}^{(22)}=\frac{k-\Omega +i\frac{1}{2}(\bar{V}_{1}^{2}-\bar{V}_{2}^{2})}{%
k-\Omega +i\frac{1}{2}(\bar{V}_{1}^{2}+\bar{V}_{2}^{2})}\,.
\end{equation}

\subsection{The state of the outgoing photons}

\begin{figure}[tbp]
\includegraphics[bb=9 2 358 234, width=7 cm, clip]{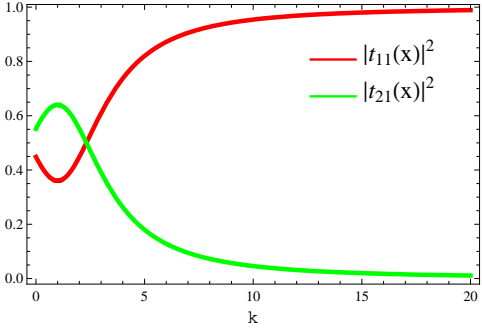}
\caption{(Color online) The transmission coefficients in the waveguide-1 and
the waveguide-2: $\bar{V}_{1}=1$, $\bar{V}_{2}=2$. The energy level spacing $%
\Omega $ is taken as the unit.}
\label{fig11}
\end{figure}

\begin{figure}[tbp]
\includegraphics[bb=9 0 358 234, width=7 cm, clip]{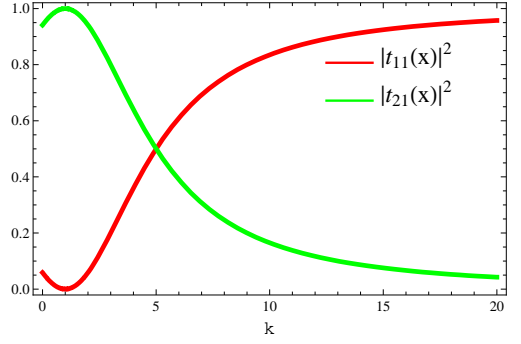}
\caption{(Color online) The transmission coefficients in the waveguide-1 and
the waveguide-2: $\bar{V}_{1}=\bar{V}_{2}=2$. The energy level spacing $%
\Omega $ of the atom is taken as the unit. If the incident photon is on
resonance with the atom, the transmission coefficient in the waveguide-1
equals zero and that in the waveguide-2 equals one. this result displays
that the outgoing photon only emits from waveguide-2.}
\label{fig12}
\end{figure}

With the help of the $S$-matrix elements obtained in the above subsection,
we find that for the single incident $e$-photon prepared in the waveguide-1,
the out-state of the scattered photon is%
\begin{equation}
\left\vert out\right\rangle =(t_{k}^{(11)}a_{k,e}^{(1)\dagger
}+t_{k}^{(21)}a_{k,e}^{(2)\dagger })\left\vert 0\right\rangle \,.
\end{equation}%
As the functions of the incident momentum $k$, the $t_{k}^{(11)}$ and $%
t_{k}^{(21)}$ are plotted in Fig.~\ref{fig11}.
We find that if $\bar{V}_{2}^{2}=\bar{V}_{1}^{2}$, the transmission
coefficient $t_{k}^{(11)}$ equals to zero when $k=\Omega $ (Fig.~\ref{fig12}%
). This result displays that the outgoing $e$-photon is only emitted from
the waveguide-2 when the incident $e$-photon is on resonance with the atom.

For the case of the two incident $e$-photons prepared in the different
waveguides, i.e., the state $\left\vert in\right\rangle $ of the incident
photons is%
\begin{equation}
\left\vert in\right\rangle =a_{k_{1},e}^{(1)\dagger }a_{k_{2},e}^{(2)\dagger
}\left\vert 0\right\rangle \,.
\end{equation}%
From Eqs. (\ref{S1112}-\ref{S2212}), we obtain the out-state of the
scattered photons as%
\begin{equation}
\left\vert out\right\rangle =\left\vert out\right\rangle _{11}+\left\vert
out\right\rangle _{12}+\left\vert out\right\rangle _{12}\,,
\end{equation}%
with three parts: (a) the state of the two outgoing $e$-photons both in the
waveguide-1 is%
\begin{equation}
\left\vert out\right\rangle _{11}=\sum_{p_{1}\leq
p_{2}}S_{p_{1}p_{2};k_{1}k_{2}}^{1,1;1,2}a_{p_{1},e}^{(1)\dagger
}a_{p_{2},e}^{(1)\dagger }\left\vert 0\right\rangle \,;
\end{equation}%
(b) the state of the two outgoing $e$-photons in the different waveguides is%
\begin{equation}
\left\vert out\right\rangle
_{12}=\sum_{p_{1},p_{2}}S_{p_{1}p_{2};k_{1}k_{2}}^{1,2;1,2}a_{p_{1},e}^{(1)%
\dagger }a_{p_{2},e}^{(2)\dagger }\left\vert 0\right\rangle \,;
\end{equation}%
(c) the state of the two outgoing $e$-photons both in the waveguide-2 is%
\begin{equation}
\left\vert out\right\rangle _{22}=\sum_{p_{1}\leq
p_{2}}S_{p_{1}p_{2};k_{1}k_{2}}^{2,2;1,2}a_{p_{1},e}^{(2)\dagger
}a_{p_{2},e}^{(2)\dagger }\left\vert 0\right\rangle \,.
\end{equation}

In the spatial representation, the state of the outgoing $e$-photon%
\begin{eqnarray}
\left\vert out\right\rangle &=&\int
dx_{1}dx_{2}e^{iEx_{c}}[g_{11}(x_{c},x)a_{x_{1},e}^{(1)\dagger
}a_{x_{2},e}^{(1)\dagger }  \notag \\
&&+g_{12}(x_{c},x)a_{x_{1},e}^{(1)\dagger }a_{x_{2},e}^{(2)\dagger }  \notag
\\
&&+g_{22}(x_{c},x)a_{x_{1},e}^{(2)\dagger }a_{x_{2},e}^{(2)\dagger
}]\left\vert 0\right\rangle \,,  \label{outreal}
\end{eqnarray}%
is determined by the wavefunction in the center of mass frame $%
x_{c}=(x_{1}+x_{2})/2$ with the relative coordinate $x=x_{1}-x_{2}$: (a) the
wavefunction $g_{11}(x)$ of two photons both in the waveguide-1 is obtained
as%
\begin{eqnarray}
g_{11}(x) &=&\frac{1}{2\pi }[t_{k_{1}}^{(11)}t_{k_{2}}^{(21)}\cos (\Delta
_{k}x)  \notag \\
&&-\frac{4\bar{V}_{2}\bar{V}_{1}^{3}e^{i(E/2-\alpha _{H})\left\vert
x\right\vert }}{4\Delta _{k}^{2}-(E-2\alpha _{H})^{2}}]\,;
\end{eqnarray}%
(b) the wavefunction $g_{12}(x)$ of two photons in the different waveguides
is obtained as%
\begin{eqnarray}
g_{12}(x) &=&\frac{1}{2\pi }%
[(t_{k_{1}}^{(11)}t_{k_{2}}^{(22)}+t_{k_{1}}^{(21)}t_{k_{2}}^{(21)})\cos
(\Delta _{k}x)  \notag \\
&&+i(t_{k_{1}}^{(11)}t_{k_{2}}^{(22)}-t_{k_{1}}^{(21)}t_{k_{2}}^{(21)})\sin
(\Delta _{k}x)  \notag \\
&&-\frac{8\bar{V}_{1}^{2}\bar{V}_{2}^{2}e^{i(E/2-\alpha _{H})\left\vert
x\right\vert }}{4\Delta _{k}^{2}-(E-2\alpha _{H})^{2}}];
\end{eqnarray}%
(c) the wavefunction $g_{22}(x)$ of the two photons both in the waveguide-2
is obtained as%
\begin{eqnarray}
g_{22}(x) &=&\frac{1}{2\pi }[t_{k_{1}}^{(21)}t_{k_{2}}^{(22)}\cos (\Delta
_{k}x)  \notag \\
&&-\frac{4\bar{V}_{1}\bar{V}_{2}^{3}e^{i(E/2-\alpha _{H})\left\vert
x\right\vert }}{4\Delta _{k}^{2}-(E-2\alpha _{H})^{2}}]\,,
\end{eqnarray}%
where we define $E=k_{2}+k_{1}$ and $\Delta _{k}=(k_{1}-k_{2})/2$.

\subsection{Quantum statistics by second order correlation functions}

\begin{figure}[tbp]
\includegraphics[bb=0 390 599 642, width=8.7 cm, clip]{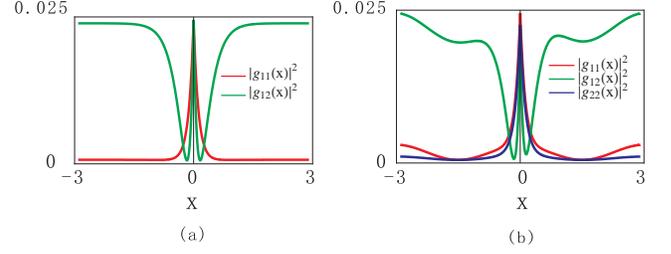}
\caption{(Color online) The correlations of the two photons in the center of
mass frame: $\bar{V}_{1}=\bar{V}_{2}=2$. The energy level spacing $\Omega $
of the atom is taken as the unit. (a) The total energy of the incident
photons $E$ is 2, and the difference $\Delta _{k}$ of two incident energies
is zero. In this case, the two photons are both on resonance on the atom.
(b) The total energy of the incident photons $E$ is 2, and the difference $%
\Delta _{k}$ of two incident energies is 1. }
\label{fig13}
\end{figure}

\begin{figure}[tbp]
\includegraphics[bb=0 390 591 636, width=8.7 cm, clip]{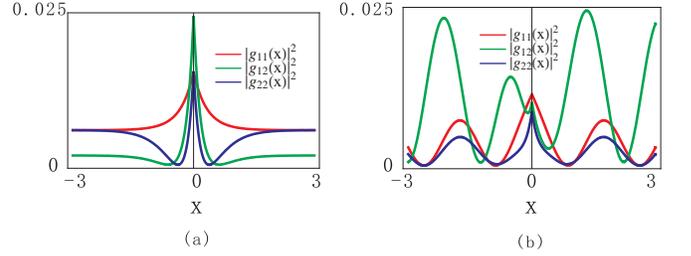}
\caption{(Color online) The correlations of the two photons in the center of
mass frame: $\bar{V}_{1}=1$ and $\bar{V}_{2}=2$. The energy level spacing $%
\Omega $ of the atom is taken as the unit. (a) The total energy of the
incident photons $E$ is 2, and the difference $\Delta _{k}$ of two incident
energies is zero. In this case, the two photons are both on resonance on the
atom. (b) The total energy of the incident photons $E$ is 2, and the
difference $\Delta _{k}$ of two incident energies is 1.8.}
\label{fig14}
\end{figure}

Finally, we analyze the quantum statistical features of the scattered photon
by the second order correlation functions of photons. The second order
correlation function of the outgoing $e$-photon is defined by%
\begin{equation}
G_{ij}^{(2)}(x_{1},x_{2})=\left\langle out\right\vert
a_{x_{1},e}^{(i)\dagger }a_{x_{2},e}^{(j)\dagger
}a_{x_{2},e}^{(j)}a_{x_{1},e}^{(i)}\left\vert out\right\rangle \,.
\label{G2}
\end{equation}%
It is straightforward to prove that the second order correlation is just $%
\left\vert g_{ij}(x)\right\vert ^{2}$ by substituting Eq. (\ref{outreal})
into Eq. (\ref{G2}). Thus the wavefunction $g_{ij}(x)$ displays the quantum
statistic characters of the scattered photons. Here, the second order
correlation function $\left\vert g_{ij}(x)\right\vert ^{2}$ are plotted in
Fig.~\ref{fig13} and Fig.~\ref{fig14} for different system parameters. In
these figures, we take the energy level spacing of the atom as unit and the
total energy of the incident two $e$-photons equal to two, i.e., two times
of the energy level spacing of the atom.

Fig.~\ref{fig13} shows that if $\bar{V}_{1}=\bar{V}_{2}=2$, the outgoing two
$e$-photons attract with each other through the interaction with the atom.
This displays the obvious bunching behavior of photons. In Fig.~\ref{fig13}%
a, the two incident photons are both on resonance with the atom, so $%
\left\vert g_{11}(x)\right\vert ^{2}=\left\vert g_{22}(x)\right\vert ^{2}$.
Fig.~\ref{fig14} shows that when $\bar{V}_{1}\neq \bar{V}_{2}$, such as $%
\bar{V}_{1}=1$ and $\bar{V}_{2}=2$, if the outgoing two $e$-photons are
emitted from the same waveguide they attract with each other and display the
photon bunching behavior. If the outgoing two $e$-photons are emitted from
the different waveguides, they display the photon bunching behavior when the
two photons are both on resonance with the atom.

By adjusting the difference $\Delta V=\bar{V}_{2}-\bar{V}_{1}$ between the
interactions $\bar{V}_{1}$ and $\bar{V}_{2}$, we find that (a) If the two
photons are both resonance with the atom, the out-going photons always
display the photon bunching behavior, however, when $\Delta V$ increases the
photon bunching behavior becomes vague and in the limit $\Delta V\rightarrow
\infty $, $\left\vert g_{11}(x)\right\vert ^{2}$, $\left\vert
g_{22}(x)\right\vert ^{2}$ and $\left\vert g_{12}(x)\right\vert ^{2}$ all
tend to constants which implies the photon bunching behavior disappears. (b)
If the two photons are not resonance with the atom and the total energy
equals to $2\Omega $, the functions $\left\vert g_{11}(x)\right\vert ^{2}$, $%
\left\vert g_{22}(x)\right\vert ^{2}$ and $\left\vert g_{12}(x)\right\vert
^{2}$ exhibit the oscillation behaviors. As $\Delta V\rightarrow \infty $, $%
\left\vert g_{11}(x)\right\vert ^{2}$, $\left\vert g_{22}(x)\right\vert ^{2}$
and $\left\vert g_{12}(x)\right\vert ^{2}$ all tend to constants and the
oscillation behavior disappears.

To summarize this section, we have used the LSZ reduction approach to
investigate the single photon transmission and the quantum statistic
properties of the two photon transport in the H-type waveguide. The
transport of multi-photon in the more complex CRA can be investigated
systematically by the same method.

\section{Summary}

In conclusion, we have demonstrated that the LSZ approach is feasible to be
generalized for studying the complex nano-structure for multi-photon
transport, such as the multi-photon transport problems in the complex CRA
systems. Concretely, the single-, two- and three- photon transports in the
T-type waveguide are investigated by the LSZ approach in details. Some of
our results accord with the known results obtained by other methods for the
simple case. Besides, the scattering of three photons in the T-type
waveguide and the quantum statistical characters of two photons in the
H-type waveguide are systematically studied by this approach. For the single
photon transmission in the T-type CRA, we find two bound states: the lower
bound state is below the bottom of the photon energy band, the upper bound
state is above the top of the photon energy band. We also gives the
three-photon scattering wave function by the LSZ approach. Obviously, the
LSZ reduction approach can be carried out to investigate the $N$-photon
transport in the more complex architectures constructed by the T-type CRA.

Finally, we emphasize that the LSZ reduction approach can be extended to
investigate not only the multi-photon scattering problem in the quantum
optics, but also the scattering processes in other fields, such as the Kondo
scattering in the condensed matter physics and the Feshbach resonance \cite%
{Xu,Feshbach} in the atomic physics. It is more interesting to simulate the
scattering processes in the condensed matter physics or the atomic physics
by the complex CRA architectures. By making use of the LSZ reduction
approach, we can study the photon scattering in these artificial CRA
architectures to understand the realistic scattering processes as well as
electron transport \cite{PALee1,PALee2,Konik} in the condensed matter
physics.

\acknowledgments This work is supported by NSFC No.~10474104, No.~60433050,
and No.~10704023, NFRPCNo.~2006CB921205 and 2005CB724508.

\appendix*

\section{Multi-photon eigenstates in the T-type CRA}

In the appendix, we utilize the scattering Bethe ansatz \cite%
{Andrei-Bethe,Andrei-Bethe-2} to obtain the multi-photon eigenstates. For
convenient, we start with the Hamiltonian (\ref{HTe}). In the real space,
the Hamiltonian (\ref{HTe}) is rewritten as $H_{T}^{e}=H_{p}+H_{a}+H_{int}$,
where the Hamiltonian of photons is%
\begin{equation}
H_{p}=-i\int dx\psi ^{\dagger }(x)\partial _{x}\psi (x)\,,
\end{equation}%
and the Hamiltonian of two- level atom is $H_{a}=\Omega \left\vert
e\right\rangle \left\langle e\right\vert $. The interaction Hamiltonian $%
H_{int}$ is%
\begin{equation}
H_{int}=\bar{V}\int dx\delta (x)(\psi (x)\sigma ^{+}+h.c.)\,.
\end{equation}%
Here,%
\begin{equation}
\psi (x)=\frac{1}{\sqrt{L}}\sum_{k}a_{k,e}e^{ikx}\,,
\end{equation}%
is the Fourier transformation of $a_{k,e}$. The single photon eigenstates
are constructed by%
\begin{equation}
\left\vert \phi _{p}\right\rangle =[\int dxf_{p}(x)\psi ^{\dagger
}(x)+e_{p}\sigma ^{+}]\left\vert 0\right\rangle \,.
\end{equation}%
The Schrodinger equation $H_{T}^{e}\left\vert \phi _{p}\right\rangle
=p\left\vert \phi _{p}\right\rangle $ gives%
\begin{equation}
\left\vert p\right\rangle =\int dxe^{ipx}\alpha _{p}^{\dagger }(x)\left\vert
0\right\rangle \,.
\end{equation}%
where we define an operator%
\begin{equation}
\alpha _{p}^{\dagger }(x)=[\theta (-x)+e^{i\delta _{p}}\theta (x)]\psi
^{\dagger }(x)+\delta (x)e_{p}\sigma ^{+}\,,
\end{equation}%
which is the single photon creation operator $\psi ^{\dagger }(x)$ when $%
\bar{V}$ tends to zero. However, when $\bar{V}$ is not zero, $\alpha
_{p}^{\dagger }(x)$ neither satisfies the bosonic commutation relation nor
the fermionic commutation relation. Here, $e_{p}$ and $f_{p}(x)$ are%
\begin{eqnarray}
e_{p} &=&\frac{V}{p-\Omega +i\Gamma /2}\,, \\
f_{p}(x) &=&e^{ipx}[\theta (-x)+e^{i\delta _{p}}\theta (x)]\,,
\end{eqnarray}%
where $\Gamma =\bar{V}^{2}$ and the phase shift%
\begin{equation}
e^{i\delta _{p}}=\frac{p-\Omega -i\Gamma /2}{p-\Omega +i\Gamma /2}\,,
\end{equation}%
is the same as $t_{k}$ (\ref{tk}).

For studying the multi-photon eigenstates, we use the scattering
Bethe-ansatz to assume the $N$-photon eigenstate%
\begin{equation}
\left\vert \Phi \right\rangle _{N}=\sum_{P}A_{P}\int
[Dx]e^{i\sum_{j}k_{P_{j}}x_{j}}\prod\limits_{i=1}^{N}\alpha
_{k_{P_{i}}}^{\dagger }(x_{i})\left\vert 0\right\rangle \,.
\end{equation}%
Here, $[Dx]$ denotes $\prod\nolimits_{i=1}^{N}\theta (x_{i+1}-x_{i})dx_{i}$
and $P=(P_{1},P_{2},...,P_{N})$ is a permutation of ($1,2,...,N$). The
Schrodinger equation gives the relation $A_{P}/A_{P^{\prime }}=e^{i\Phi
(P_{j},P_{j+1})}$, where the phase shift is%
\begin{equation}
e^{i\Phi (P_{j},P_{j+1})}=\frac{k_{P_{j}}-k_{P_{j+1}}-i\Gamma }{%
k_{P_{j}}-k_{P_{j+1}}+i\Gamma }\,,
\end{equation}%
where $P=(P_{1},P_{2},...,P_{j},P_{j+1},...,P_{N})$ and $P^{\prime
}=(P_{1},P_{2},...,P_{j+1},P_{j},...,P_{N})$. As an example, we give the
explicit expression of the three photon eigenstate%
\begin{eqnarray}
\left\vert \Phi \right\rangle _{N=3} &=&\int \theta (x_{3}-x_{2})\theta
(x_{2}-x_{1})\prod\limits_{i}dx_{i}  \notag \\
&&[1+S_{12}+S_{23}+S_{12}S_{13}  \notag \\
&&+S_{23}S_{13}+S_{12}S_{13}S_{23}]e^{ik_{1}x_{1}+ik_{2}x_{2}+ik_{3}x_{3}}
\notag \\
&&\times \alpha _{k_{1}}^{\dagger }(x_{1})\alpha _{k_{2}}^{\dagger
}(x_{2})\alpha _{k_{3}}^{\dagger }(x_{3})\left\vert 0\right\rangle \,,
\end{eqnarray}%
where%
\begin{equation}
S_{ij}=\frac{k_{j}-k_{i}-i\Gamma }{k_{j}-k_{i}+i\Gamma }P_{ij}\,,
\end{equation}%
and $P_{ij}f(...,k_{i},...,k_{j},...)=f(...,k_{j},...,k_{i},...)$. Here, $f$
is any given function.


\begin{thebibliography}{99}
\bibitem{NP07} D. E. Chang, A. S. S\o rensen, E. A. Demler, and M. D. Lukin,
Nat. Phys. \textbf{3}, 807 (2007).

\bibitem{Fan05} J. T. Shen and S. Fan, Phys. Rev. Lett. \textbf{95}, 213001
(2005).

\bibitem{Sun} C. P. Sun, L. F. Wei, Yu-xi Liu, and F. Nori,\textit{\ }Phys.
Rev. A \textbf{73}, 022318 (2006).

\bibitem{Zhou} L. Zhou, Z. R. Gong, Y. X. Liu, C. P. Sun and F. Nori, Phys.
Rev. Lett. \textbf{101}, 100501 (2008).

\bibitem{ST} T. Shi, Y. Li, Z. Song, and C. P. Sun, Phys. Rev. A \textbf{71}%
, 032309 (2005).

\bibitem{YS1} S. Yang, Z. Song, and C. P. Sun, Phys. Rev. B \textbf{73},
195122 (2006).

\bibitem{YS2} S. Yang, Z. Song, and C. P. Sun, Front. Phys. China \textbf{2}%
, 1 (2007).

\bibitem{FanPRL} J. T. Shen and S. Fan, Phys. Rev. Lett. \textbf{98}, 153003
(2007).

\bibitem{FanPRA} J. T. Shen and S. Fan, Phys. Rev. A \textbf{76}, 062709
(2007).

\bibitem{FanOpt} J. T. Shen and S. Fan, Opt. Lett. \textbf{30}, 2001 (2005).

\bibitem{Zhou2} L. Zhou, J. Lu, and C. P. Sun, Phys. Pev. A \textbf{76},
012313 (2007).

\bibitem{FMH} F. M. Hu, L. Zhou, T. Shi, and C. P. Sun, Phys. Rev. A \textbf{%
76}, 013819 (2007).

\bibitem{Nori2} A. L. Rakhmanov, A. M. Zagoskin, S. Savel'ev, and F. Nori,
Phys. Rev. B \textbf{77}, 144507 (2008).

\bibitem{Dicke} R. H. Dicke, Phys. Rev. \textbf{93}, 99 (1954).

\bibitem{Nori} K. Y. Bliokh, Y. P. Bliokh, V. Freilikher, S. Savel'ev, and
F. Nori, Rev. Mod. Phys. \textbf{80}, 1201 (2008).

\bibitem{F} U. Fano, Phys. Rev. \textbf{124}, 1866 (1961).

\bibitem{A} P. W. Anderson, Phys. Rev. \textbf{124}, 41 (1961).

\bibitem{L} T. D. Lee, Phys. Rev. \textbf{95}, 1329 (1954).

\bibitem{Xu} D. Z. Xu, H. Ian, T. Shi, H. Dong, and C. P. Sun, arXiv:
quant-ph/0812.0429.

\bibitem{Landau} L. D. Landau and E. M. Lifshitz, \textit{Quantum Mechanics
(Nonrelativistic Theory)} (Butterworth, Boston 1991).

\bibitem{Bethe} H. A. Bethe, Z. Physik \textbf{71}, 205 (1931).

\bibitem{Yang} C. N. Yang, Phys. Rev. Lett. \textbf{19}, 1312 (1967).

\bibitem{Batchelor} M. T. Batchelor, Phys. Today \textbf{60}, 36 (2007).

\bibitem{KondoAndrei} N. Andrei, Phys. Rev. Lett. \textbf{45}, 379 (1980).

\bibitem{Wiegmann1} P. B. Wiegmann, JETP Lett. \textbf{31}, 364 (1980).

\bibitem{Wiegmann2} P. B. Wiegmann, J. Phys. C \textbf{14}, 1463 (1981).

\bibitem{Wiegmann3} P. B. Wiegmann and A. M. Tsvelick, J. Phys. C \textbf{16}%
, 2281 (1983).

\bibitem{LSZ} H. Lehmann, K. Symanzik, and W. Zimmermann, Nuovo Cimento
\textbf{1}, 1425 (1955).

\bibitem{Andrei1} G. Zarand, L. Borda, J. Delft, and N. Andrei, Phys. Rev.
Lett. \textbf{93}, 107204 (2004).

\bibitem{Andrei2} L. Borda, L. Fritz, N. Andrei, and G. Zarand, Phys. Rev. B
\textbf{75}, 235112 (2007).

\bibitem{Pauli} G. K\"{a}ll\'{e}n and W. Pauli, Dan., Mat. Fys. Medd.
\textbf{30}, 7 (1955).

\bibitem{Lee-Wick} T. D. Lee and G. C. Wick, Nucl. Phys. B \textbf{10}, 1
(1969).

\bibitem{Maxon1} M. S. Maxon and R. B. Curtis, Phys. Rev. \textbf{137}, B996
(1965).

\bibitem{Maxon2} M. S. Maxon, Phys. Rev. \textbf{149}, 1273 (1966).

\bibitem{Sakurai} J. J. Sakurai, \textit{Moden Quantum Mechanics}
(Addison-Wesley, Reading, MA, 1994).

\bibitem{Taylor} J. R. Taylor, \textit{Scattering Theory: The Quantum Theory
on Non-relativistic Collision} (Wiley, New York, 1972).

\bibitem{Andrei-Bethe} P. Mehta and N. Andrei, Phys. Rev. Lett. \textbf{96},
216802 (2006).

\bibitem{Andrei-Bethe-2} P. Mehta and N. Andrei, arXiv:cond-mat/0702612.

\bibitem{Feshbach} H. Feshbach, Ann. Phys. \textbf{5}, 357-390 (1958).

\bibitem{PALee1} T. K. Ng and P. A. Lee, Phys. Rev. Lett. \textbf{61}, 1768
(1988).

\bibitem{PALee2} Y. Meir, N. S. Wingreen, and P. A. Lee, Phys. Rev. Lett.
\textbf{70}, 2601 (1993).

\bibitem{Konik} R. M. Konik, H. Saleur, and A. Ludwig, Phys. Rev. B \textbf{%
66}, 125304 (2002).
\end{thebibliography}
\end{document}